\documentclass[twocolumn,prb,10pt,aps,longbibliography,superscriptaddress]{revtex4-2}

\usepackage{graphicx}
\usepackage{grffile}
\usepackage{amsmath}
\usepackage{amssymb}
\usepackage{amsfonts}
\usepackage{dsfont}
\usepackage{dcolumn}
\usepackage{bbm}
\usepackage{subfigure}
\usepackage{mathtools}
\usepackage{braket}
\usepackage{placeins}
\usepackage{physics}
\usepackage{longtable}
\usepackage{siunitx}
\usepackage[unicode]{hyperref}
\usepackage[nameinlink]{cleveref}
\Crefname{section}{Sec.}{Secs.}
\Crefname{equation}{Eq.}{Eqs.}
\Crefname{figure}{Fig.}{Figs.}
\Crefname{tabular}{Tab.}{Tabs.}

\usepackage{bbold} % For \mathbb{1}
\usepackage{nicefrac}
\usepackage{tikz}

\newcommand{\tjm}[0]{$t$--$J$ model}
\newcommand{\Uc}{U_\mathrm{c}{}} % Achtung, geaendert
\newcommand{\ieom}[0]{\text{iEoM}}
\newcommand{\cet}[0]{\text{CET}}

\newcommand{\Liouville}[1]{\mathcal{L}\!\left(#1\right)}

\newcommand{\normfactor}[0]{\mathcal{N}}
\newcommand{\matrixsymbol}[1]{\mathbf{#1}}
\renewcommand\vec{\mathbf}
\newcommand{\scalarpr}[2]{\left(#1\middle|#2\right)}

\usepackage{xstring}
\newcommand{\refcite}[1]{%
\begingroup
\def\tempx{0}%
  \StrCount{#1}{,}[\tempx]%
  \ifnum\tempx > 0 
  Refs.~%
  \else
  Ref.~%
  \fi
\endgroup
\cite{#1}%
}

\newcommand{\Dim}[0]{d} % Normfaktor beim Skalarprodukt

\newcommand{\bes}{\begin{subequations}}
\newcommand{\ees}{\end{subequations}}

\newcommand{\erz}[1]{f_{#1}^\dagger}

\newcommand{\herz}[1]{h_{#1}^\dagger}

\newcommand{\ver}[1]{f_{#1}^{\vphantom{\dagger}}}

\newcommand{\bzo}[1]{\widehat{n}_{#1}}

\NewDocumentCommand\crea{g} {\ensuremath{f^{\dagger}_{   \IfNoValueTF{#1}{}{ #1} } }}
\NewDocumentCommand\anhi{g} {\ensuremath{f^{\phantom{\dagger}}_{   \IfNoValueTF{#1}{}{ #1} } } }
\NewDocumentCommand\dtup{ s g }{ \ensuremath{ \IfBooleanTF#1
{\crea{  \IfNoValueTF{#2}{\uparrow}{ #2, \uparrow }    } \crea{ \IfNoValueTF{#2}{ \downarrow}{ #2, \downarrow} } \anhi{ \IfNoValueTF{#2}{ \downarrow}{ #2, \downarrow} } } 
{\overline{d}^{\dagger}_{   \IfNoValueTF{#2}{\uparrow}{ #2, \uparrow } } } } }
\NewDocumentCommand\dtdown{ s g }{ \ensuremath{ \IfBooleanTF#1
{\crea{  \IfNoValueTF{#2}{\downarrow}{ #2, \downarrow }    } \crea{ \IfNoValueTF{#2}{ \uparrow}{ #2, \uparrow} } \anhi{ \IfNoValueTF{#2}{ \uparrow}{ #2, \uparrow} } }   
{\overline{d}^{\dagger}_{   \IfNoValueTF{#2}{\downarrow}{ #2, \downarrow } } } }}
\NewDocumentCommand\dup{ s g }{ \ensuremath{\IfBooleanTF#1
{\anhi{  \IfNoValueTF{#2}{\uparrow}{ #2, \uparrow }    } \crea{ \IfNoValueTF{#2}{ \downarrow}{ #2, \downarrow} } \anhi{ \IfNoValueTF{#2}{ \downarrow}{ #2, \downarrow} } } 
{\overline{d}^{\phantom{\dagger}}_{   \IfNoValueTF{#2}{\uparrow}{ #2, \uparrow } } } } }
    
\NewDocumentCommand\ddown{ s g }{ \ensuremath{\IfBooleanTF#1
{\anhi{  \IfNoValueTF{#2}{\downarrow}{ #2, \downarrow }    } \crea{ \IfNoValueTF{#2}{ \uparrow}{ #2, \uparrow} } \anhi{ \IfNoValueTF{#2}{ \uparrow}{ #2, \uparrow} } }   
{\overline{d}^{\phantom{\dagger}}_{   \IfNoValueTF{#2}{\downarrow}{ #2, \downarrow } } } }}
\NewDocumentCommand\ltup{ s g }{ \ensuremath{ \IfBooleanTF#1
{\anhi{  \IfNoValueTF{#2}{\uparrow}{ #2, \uparrow }    } \anhi{ \IfNoValueTF{#2}{ \downarrow}{ #2, \downarrow} } \crea{ \IfNoValueTF{#2}{ \downarrow}{ #2, \downarrow} } } 
{\overline{h}^{\dagger}_{   \IfNoValueTF{#2}{\uparrow}{ #2, \uparrow } } } } }
\NewDocumentCommand\ltdown{ s g }{ \ensuremath{ \IfBooleanTF#1
{\anhi{  \IfNoValueTF{#2}{\downarrow}{ #2, \downarrow }    } \anhi{ \IfNoValueTF{#2}{ \uparrow}{ #2, \uparrow} } \crea{ \IfNoValueTF{#2}{ \uparrow}{ #2, \uparrow} } }   
{\overline{h}^{\dagger}_{   \IfNoValueTF{#2}{\downarrow}{ #2, \downarrow } } } }}
\NewDocumentCommand\lup{ s g }{ \ensuremath{\IfBooleanTF#1
{\crea{  \IfNoValueTF{#2}{\uparrow}{ #2, \uparrow }    }  \anhi{ \IfNoValueTF{#2}{ \downarrow}{ #2, \downarrow} } \crea{ \IfNoValueTF{#2}{ \downarrow}{ #2, \downarrow} } } 
{\overline{h}^{\phantom{\dagger}}_{   \IfNoValueTF{#2}{\uparrow}{ #2, \uparrow } } } } }
    
\NewDocumentCommand\ldown{ s g }{ \ensuremath{\IfBooleanTF#1
{\crea{  \IfNoValueTF{#2}{\downarrow}{ #2, \downarrow }    } \anhi{ \IfNoValueTF{#2}{ \uparrow}{ #2, \uparrow} } \crea{ \IfNoValueTF{#2}{ \uparrow}{ #2, \uparrow} } }   
{\overline{h}^{\phantom{\dagger}}_{   \IfNoValueTF{#2}{\downarrow}{ #2, \downarrow } } } }}
\NewDocumentCommand\splus{ s g }{ \ensuremath{ \IfBooleanTF#1
{ \crea{  \IfNoValueTF{#2}{\uparrow}{ #2, \uparrow }    } \anhi{ \IfNoValueTF{#2}{ \downarrow}{ #2, \downarrow} }   }
{\sigma^{+}_{\IfNoValueTF{#2}{}{#2} } }      } }
\NewDocumentCommand\sminus{ s g }{ \ensuremath{ \IfBooleanTF#1
{ \crea{  \IfNoValueTF{#2}{\downarrow}{ #2, \downarrow }    } \anhi{ \IfNoValueTF{#2}{ \uparrow}{ #2, \uparrow} }   } 
{\sigma^{-}_{\IfNoValueTF{#2}{}{#2} } }  } }
\NewDocumentCommand\sz{ s g }{ \ensuremath{ \IfBooleanTF#1
{ n^{\phantom{\dagger}}_{ \IfNoValueTF{#2}{\uparrow}{ #2, \uparrow }    } - n^{\phantom{\dagger}}_{ \IfNoValueTF{#2}{ \downarrow}{ #2, \downarrow} }   }  
{\sigma^{z}_{\IfNoValueTF{#2}{}{#2} } }    } }
\NewDocumentCommand\nbar{ s g }{ \ensuremath{ \IfBooleanTF#1
{ n^{\phantom{\dagger}}_{ \IfNoValueTF{#2}{\uparrow}{ #2, \uparrow }    } + n^{\phantom{\dagger}}_{ \IfNoValueTF{#2}{ \downarrow}{ #2, \downarrow} } - \mathbb{1}  }     
{\bar{n}^{\phantom{\dagger}}_{\IfNoValueTF{#2}{}{#2} } } } }
\NewDocumentCommand\cc{g}{ \ensuremath{  \anhi{  \IfNoValueTF{#1}{\downarrow}{ #1, \downarrow }    } 
\anhi{ \IfNoValueTF{#1}{ \uparrow}{ #1, \uparrow} }   } }
\NewDocumentCommand\ctct{g}{ \ensuremath{  \crea{  \IfNoValueTF{#1}{\downarrow}{ #1, \downarrow }    } 
\crea{ \IfNoValueTF{#1}{ \uparrow}{ #1, \uparrow} }   } }
\NewDocumentCommand\nU{ s g }{ \ensuremath{ \IfBooleanTF#1{
 \left( n^{\phantom{\dagger}}_{ \IfNoValueTF{#2}{ \downarrow}{ #2, \downarrow} } - \frac{1}{2} \right)        
 \left( n^{\phantom{\dagger}}_{ \IfNoValueTF{#2}{\uparrow}{ #2, \uparrow } } - \frac{1}{2} \right) 
}   
{n^{u}_{\IfNoValueTF{#2}{}{#2} } } } } 
\NewDocumentCommand\nup{g }{ \ensuremath{\left(n^{\phantom{\dagger}}_{   \IfNoValueTF{#1}{\uparrow}{ #1, \uparrow } }- \frac{1}{2} \right) } }
\NewDocumentCommand\ndown{g }{ \ensuremath{\left(n^{\phantom{\dagger}}_{   \IfNoValueTF{#1}{\downarrow}{ #1, \downarrow } }- \frac{1}{2} \right) } }
\NewDocumentCommand\n{s g}{ \ensuremath{ \IfBooleanTF#1
{\crea{#2} \anhi{#2}}
{\hat{n}^{\phantom{\dagger}}_{   \IfNoValueTF{#2}{}{ #2} } } } }
\NewDocumentCommand\ntilde{g}{ \ensuremath{ 
\tilde{n}^{\phantom{\dagger}}_{   \IfNoValueTF{#1}{}{ #1} } } } 
\NewDocumentCommand\gcite{g} {Ref.~\cite{#1}}
\NewDocumentCommand\Liou{g} {\ensuremath{\mathcal{L}\left( #1 \right)     }   }
\NewDocumentCommand\fnorm{g g} {\ensuremath{  ( #1  \vert #2 )     }   }
\NewDocumentCommand\vect{g} {\ensuremath{  \bm{#1}  } }
\NewDocumentCommand\mat{g} {\ensuremath{  \text{\textbf{#1}}  } }

\begin{document}

\title{Charge dynamics in magnetically disordered Mott insulators}

\author{Philip Bleicker}
\email{philip.bleicker@tu-dortmund.de}
\affiliation{Lehrstuhl f\"{u}r Theoretische Physik I, 
Technische Universit\"{a}t Dortmund,
 Otto-Hahn-Stra\ss{}e 4, 44221 Dortmund, Germany}

\author{Dag-Bj\"orn Hering}
\email{dag.hering@tu-dortmund.de}
\affiliation{Lehrstuhl f\"{u}r Theoretische Physik I, 
Technische Universit\"{a}t Dortmund,
 Otto-Hahn-Stra\ss{}e 4, 44221 Dortmund, Germany}

\author{G\"otz S.\ Uhrig}
\email{goetz.uhrig@tu-dortmund.de}
\affiliation{Lehrstuhl f\"{u}r Theoretische Physik I, 
Technische Universit\"{a}t Dortmund,
 Otto-Hahn-Stra\ss{}e 4, 44221 Dortmund, Germany}

\date{\textrm{\today}}

\begin{abstract}
With the aid of both a semi-analytical and a numerically exact method we 
investigate the charge dynamics 
in the vicinity of half-filling in the one- and two-dimensional 
 {$t$-$J$ model} derived from a Fermi-Hubbard model 
in the limit of large interaction $U$ and hence small exchange coupling $J$.
The spin degrees of freedom are taken to be disordered. 
So we consider the limit $0 < J \ll T \ll W$ where $W$ is the band width.
We focus on evaluating the  {spectral density} of a single hole 
excitation and the charge gap which separates the upper and the lower 
Hubbard band.  {One of the key findings is the evidence for the absence of sharp
edges of the Hubbard band, instead Gaussian tails appear.} 
\end{abstract}

%\pacs{05.70.Ln, 67.85.−d, 71.10.Fd, 71.10.Pm}
%05.70.Ln	Nonequilibrium and irreversible thermodynamics
%67.85.−d	Ultracold gases, trapped gases
%71.10.Fd	Lattice fermion models (Hubbard model, etc.)
%71.10.Pm 	Fermions in reduced dimensions

\maketitle

\section{Introduction}
\label{s:introduction}

Strongly correlated fermionic systems and Mott-Hubbard physics
in particular continue to represent a great challenge to 
theoretical treatments in spite of many decades of research
\cite{gebha97}. Even rather
clear physical questions cannot be answered in a straightforward manner.
A prominent example is the motion of a single hole in a Mott insulator. This issue has 
attracted a lot of interest early after the discovery of high-temperature superconductivity
because it was noted that the hopping hole scrambles the antiferromagnetic background
which can act as attractive force between two holes if the second one heals
the misalignments caused by the first hole \cite{trugm88}. The motion of holes in 
ordered antiferromagnets continues to be a topic of current research \cite{bonca07},
nowadays extended also to non-equilibrium situations \cite{mierz11a}.

Equally, the hole motion in a \emph{disordered} spin background is a highly non-trivial issue.
At first glance, one may think that there is no order to be scrambled such 
that the hole can move as freely as it does without any interaction
so that the single particle Mott gap $\Delta$ is given by 
$ (W -U)/2$ where $W$ is the band width and $U$ the local Hubbard repulsion.
This expectation, however, is only correct in the extreme limit $U\to\infty$ 
and for hole motion on self-retracing paths, for instance in one dimension (1D)
\cite{mielk91aa,kumar09,Nocera2018}. For finite values of $U$ in a Hubbard model
even the infinite-dimensional case yields a non-trivial value for
the opening of the Mott gap computed to lie between $\Uc\approx 1.11W$ \cite{eastw03,Nishimoto2004}
and $\Uc\approx 1.19W$ \cite{bulla99a,bulla01a,garci04,blume05a,karsk05,karsk08}.
Note that in the considered paramagnetic infinite-dimensional case 
the spin background is indeed completely disordered without spin-spin correlations
between different sites.

In 1D, the Bethe ansatz allows for an exact treatment \cite{essle05}
showing a Mott insulator at half-filling and zero temperature
for infinitesimal interaction $U$. But
it is also possible to consider a completely disordered spin background \cite{Ejima2006}
corresponding to the situation where $J\ll T\ll U \approx W$. Here, $J$ is the
nearest-neighbor (NN) antiferromagnetic exchange coupling taking the value $4t_0^2/U$ 
in leading order in the NN hopping $t_0$ 
\cite{ander59b,harri67,klein73,takah77,macdo88,stein97} and $W$ is the band width.
Under this assumption, a Mott transition is identified to occur at 
$\Uc=\sqrt{3}W/2\approx 0.866W$. This finding provides an important benchmark.
All these results illustrate that the hole motion is influenced by non-trivial
quantum effects even for disordered spin backgrounds.

The aim of the present article  {is to study} the hole motion
in one and two dimensions (2D), i.e., along a chain and on a square lattice.
The former case serves both as a benchmark and, due to its lower coordination number, 
as a system in which a larger number of processes with a larger spread is numerically 
accessible than in lattices with higher coordination numbers. This facilitates an in-depth 
spectral analysis and, in particular, the analysis of the typically difficult-to-access 
edges of the excitation spectrum. The latter case of a two-dimensional square lattice 
actually represents the most interesting case in view of experimental realizations
in solid state systems or in cold atom setups.
We consider the $t$-$J$ model which is derived from the Hubbard model
\cite{ander59b,harri67,klein73,takah77,macdo88,eskes94b,stein97}. We stress that
the mapping from the Hubbard model to the $t$-$J$ model is not
restricted to the magnetic exchange couplings, but naturally extends
to the charge degrees of freedom, i.e., to hopping terms, hole-hole interactions,
and correlated hopping processes. This applies to the chain \cite{mielk91aa} 
and to the square lattice \cite{eskes94b,Reischl2004} at half-filling, but also in the vicinity
of half-filling, i.e., for finite doping \cite{hamer10}. 

We proceed in two steps. First, we consider very large $U$, i.e., we 
omit all terms of order $t_0^2/U$ and only keep terms of order $U$ and $t_0$.
Second, we include the terms of order $t_0^2/U$ to study to which extent
they induce changes in the spectral densities  {including the character of the
band edges. Such changes are expected, for instance the critical $U$ deviates from
$W$ in the estimate $\Uc\approx 1.10 W$ obtained by Reischl et al.\,\cite{Reischl2004}
for} the square lattice.

A  semi-analytic and a  numeric approach are employed. 
The first, semi-analytic, approach relies on iterated equations of motion (\ieom{})
in the Heisenberg picture. The set of tracked operators is
enlarged iteratively by commuting with the Hamiltonian, i.e.
by applying the Liouville superoperator. This Liouvillean 
acts on operators like a Hamiltonian acts on states 
\cite{Kalthoff2017,Bleicker2018} yielding a Hermitian, oscillatory dynamics.
The dominant part of the Liouvillean is the 
commutation with the hopping projected in such a way
that no double occupancies are created or annihilated.
Thus, the semi-analytic approach amounts up to a systematic
expansion in the hopping element, that means in $x:=t_0/U$.
The second, numeric,
approach tracks the hole motion in time on finite clusters with periodic
boundary conditions in 1D and 2D by Chebyshev
polynomial expansion \cite{Tal-EzerH.1984,weiss06a,Bleicker2020}. 

This article is structured in the following way: 
In \Cref{s:model}, the Hubbard model and its simplification in the limit of strong interaction
 is explained briefly. \Cref{s:method} outlines the concepts and algorithms
used to access the time-evolution of observables and to gain insight into the metal-insulator phase.
  {\Cref{s:method_comparison} provides data in the time domain 
comparing results from the two approaches used and illustrates 
how band edges are determined.}
 In \Cref{s:results-1D} and \Cref{s:results-2D} we discuss the results 
for the \tjm{} on the one-dimensional chain and on the two-dimensional square lattice,
respectively. Summary and outlook are given in \Cref{s:summary}.

\section{Initial Model} 
\label{s:model}

The Fermi-Hubbard model is one of the prime examples and 
archetypical models for strongly interacting electrons 
on a lattice and combines tight-binding electrons with a 
strongly screened Coulomb interaction \cite{Hubbard1963,Kanamori1963,Gutzwiller1963}. 
In the following, we restrict our considerations to the one-band model in the 
vicinity of half-filling such that the Hamiltonian takes the form
\begin{subequations}
\label{eq:fhm_real}
\begin{alignat}{3}
	H &= H_0 + H_\text{int} \\
	H_0 &= t_0 \sum_{\substack{\langle i,j\rangle, \sigma}} {\vphantom{\dagger}}
	(\erz{i\sigma} \ver{j\sigma} +\text{h.c.}) \\
	H_\text{int} &= U \sum_{i}\left(\bzo{{i\uparrow}}-\frac{1}{2}\right) \left(\bzo{{i\downarrow}}-\frac{1}{2}\right).
	\label{eq:wewi1}
\end{alignat}
\end{subequations}
Here, $\erz{i\sigma}$ ($\ver{i\sigma}$) are the creation (annihilation) operators
at site $i$ for a fermion of spin $\sigma$ and $\bzo{{i\sigma}}$ is the corresponding
number operator, $t_0$ denotes the real hopping matrix element between the sites $i$ 
and $j$ and $U>0$ is the on-site interaction. As denoted in \eqref{eq:wewi1},
$U/2$ represents the energy cost if an electron is added inducing a double occupancy (DO)
of two electrons at one site or, if a hole is added, inducing a double occupancy of holes at one site,
i.e., creating an empty site.
The kinetic energy $H_0$ is diagonal in momentum space such that all quasi-particles 
obey the dispersion relation 
\begin{equation}
	\label{eq:hubbard_dispersion_momentum}
	\varepsilon_\vec{k} := 2t \sum_{i=1}^d \cos{\left( \vec{k}\vec{a}_i \right)}
\end{equation}
with $\vec{a}_i$ denoting primitive translation vectors spanning the underlying Bravais lattice.
The model is particle-hole symmetric on bipartite lattices such as the 1D chain or the 2D square lattice.

\subsection{Charge gap}

\begin{figure}[ht]
  \centering
  \includegraphics[trim=70 10 10 0,clip,width=.5\textwidth]{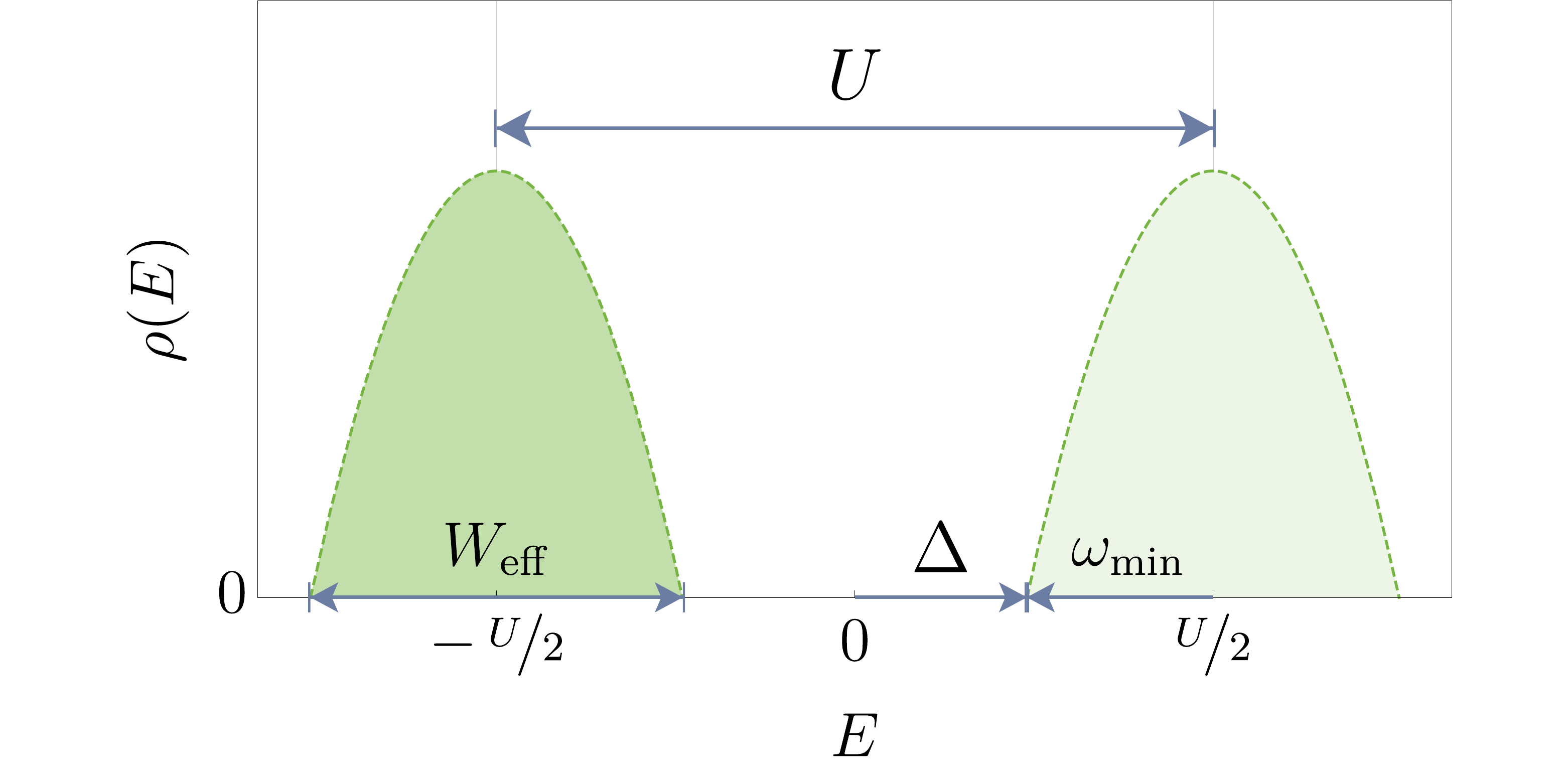}
  \caption{For a large enough $U>0$, the local density of states $\rho(E)$ splits into a lower and 
	an upper Hubbard band at half-filling. Each band has the effective band width $W_\mathrm{eff}$. 
	The LHB is completely filled, the UHB is empty. Charge excitations take the form of DOs of electron
	or hole character with a 	minimum excitation energy of the gap $\Delta$. 
	Decreasing the on-site interaction reduces the gap until it closes at the critical interaction $\Uc$.}
  \label{img:hubbard_bands_gap}
\end{figure}

The introduction of a large enough on-site interaction $U>0$ in the Hubbard model splits
 the local density-of-states $\rho(E)$ into a lower (LHB) and an upper Hubbard band (UHB) as shown in 
\Cref{img:hubbard_bands_gap}.

Except for special cases, i.e., for specific lattices, cf. Introduction \ref{s:introduction}, 
it remains an open question how the charge gap $\Delta$ behaves upon diminishing interaction strengths,
for instance at which critical interaction strengths $\Uc$ the gap finally closes
 {signalling the instability of the Mott insulator}.

In this context, the 1D model plays a special role due to its integrability. It can be
 solved exactly by means of thermodynamic Bethe ansatz equations \cite{essle05} at any
temperature, but also under the assumption of a disordered spin background and
charge excitations at zero temperature \cite{Ejima2006}.  {Yet we are not aware
that the spectral density of the Hubbard models has been determined
exactly by Bethe ansatz.} Another possibility at 
finite temperatures is to use time-dependent density matrix renormalization group 
computations  {which allows one to determine spectral properties
as well} \cite{Nocera2018}. Moreover, the complementary limit of a Bethe lattice with an 
infinitely large coordination number $z\to\infty$
can be treated by dynamic mean-field theory  {readily providing information on
the local spectral densities}
 \cite{bulla99a,bulla01a,eastw03,Nishimoto2004,garci04,blume05a,karsk05,karsk08}.
Otherwise, statements regarding the dynamics of charge carriers are difficult to provide. 
We address this very question using two fundamentally different techniques in the remainder of the work.

\subsection{Effective Model:  {$t$-$J$ Model}}
\label{ss:effective_model}

In the limit of strong interaction, i.e., $x=\nicefrac{t_0}{U}\ll1$, the Hubbard model can be 
 {mapped to the $t$-$J$ model and thereby simplify it
 based on perturbation theory in the small parameter $x$}. We consider all  {linear $U$} 
 terms in zeroth order as well as the $xU=t_0$ term $H_0$
 describing hopping in first order. 
Contributions of second order, i.e., especially magnetic superexchange $J\propto x^2 U$, 
and contributions of even higher orders are neglected in the first step,  {but will
be considered in a second step. First, we study} a Hamiltonian of the general form
\begin{equation}
	H_\mathrm{eff} = H_\text{int} + H_0 + \mathcal{O}\left(x^2 U\right).
\end{equation}
A  systematic approach to derive such an effective model is to resort to a 
continuous unitary transformation (CUT) 
\cite{Wegner1994,stein97,mielk98,knett00a,Reischl2004,kehre06,hamer10}. 
For the sake of completeness, we will briefly recall the concepts as presented in 
\refcite{Reischl2004} here. For a CUT one conventionally starts with the flow equation 
\cite{Wegner1994} as given by
\begin{equation}
  \label{eq:flow_equation}
  \frac{d}{d\ell}{H}(\ell)=\comm{\eta(\ell)}{H(\ell)}
\end{equation}
with a properly chosen antihermitian generator $\eta(\ell)$. Similar to the perturbative
reasoning \cite{ander59b,harri67,klein73,takah77,macdo88}, the key idea is to 
eliminate processes which \emph{change} the number of 
DOs so that the final effective Hamiltonian preserves the number of DOs.
To this end, one can choose 
\begin{equation}
	\eta(\ell)=\comm{\hat D}{H(\ell)}
\end{equation}
where $\hat D :=\sum_i [n_{i,\uparrow}n_{i,\downarrow} + (1-n_{i,\uparrow})(1-n_{i,\downarrow})]$ 
counts the number of DOs, i.e., all sites occupied by either two particles or completely empty.
We stress that the resulting generator is proportional to the ones with \emph{sign} functions
\cite{mielk98,knett00a,Reischl2004} due to the simplicity of $\hat D$.
 Governed by the flow equation \eqref{eq:flow_equation} a transformation from the initial 
Hamiltonian $H(0)=H$ to the effective Hamiltonian 
\begin{equation}
	H_\mathrm{eff}=\lim_{\ell\to\infty}H(\ell)	
\end{equation}
takes place. It is this effective Hamiltonian one is usually interested in. Obviously, it is 
necessary to restrict the number of contributing operator terms
generated by the flow equation in a suitable manner. One possibility, employed
 in \refcite{Reischl2004,hamer10}, is to define a proper measure of locality: operator terms which
are not sufficiently local are discarded. Since the non-locality of the Hubbard model
is due to the hopping, this approach corresponds to an expansion in $t_0$.
This procedure eventually leads to a generalized \tjm{} at half-filling
\cite{Reischl2004} and for moderate doping \cite{hamer10}. The leading
order at half-filling is known also analytically \cite{mielk91aa,eskes94b,kumar09}.

Exploiting the assumption $0 < J \ll T \ll W$ where the temperature $T$ is much larger 
than the typical magnetic coupling strength $J$ we consider a  {completely} disordered spin background. The charge carriers, however, behave as if the system were essentially at vanishing absolute temperature $T\approx0$. 
The effective hopping processes in the \tjm{}  {can be split in the following way}
\begin{equation}
    \label{eq:generalized_tj_model}
	 H_\mathrm{0,eff} = T_0 + T'_0 + T'_{s,0} + T''_0 + T''_{s,0}.
\end{equation}
For the sake of brevity, we will use the term \tjm{} in the following for the above hopping  {model}
even if the magnetic exchange is not  {present}. The magnetic exchange interaction can easily be  {added}
to $H_\mathrm{0,eff}$. We will  {come  to this complete \tjm{} at a later stage of this article}.

In \Cref{eq:generalized_tj_model} the term $T_0$ describes 
nearest-neighbor (NN) hopping from site $i$ to $j$ and vice-versa 
subjected to  the restraint that DOs neither are added nor removed, i.e.,
\begin{eqnarray}
T_0&=& t_0\sum_{\mathclap{\langle i,j\rangle, \sigma}}
  \Big[(1-n^{\phantom{\dagger}}_{i,\sigma}) c^\dagger_{i,\overline{\sigma}}
  c^{\phantom{\dagger}}_{j,\overline{\sigma}} 
(1-n^{\phantom{\dagger}}_{j,\sigma})\nonumber \\
\label{eq:T0}
&&\phantom{t_0\sum_{{ \langle i,j\rangle, \sigma}}\Big[}
  +n^{\phantom{\dagger}}_{i,\sigma} c^\dagger_{i,\overline{\sigma}}
   c^{\phantom{\dagger}}_{j,\overline{\sigma}} 
n^{\phantom{\dagger}}_{j,\sigma}+\text{h.c.}\Big].
\end{eqnarray}
Here and in the following, the notation  {of the sums} means a one-time counting of 
each bond between the lattice sites $i$ and $j$ and $\overline{\sigma}$ denotes the opposite of the orientation $\sigma$. The generalization of such hopping 
to next-nearest neighbor (NNN) hopping processes, i.e., 
all processes between sites on the 2D square lattice which lie on adjacent \emph{diagonal}
 positions, are denoted by
\begin{eqnarray}
\label{eq:NNN_hoppings}
  T'_0&=& t'\sum_{\mathclap{\langle \langle i,j\rangle\rangle; \sigma}}
  \Big[(1-n^{\phantom{\dagger}}_{i,\sigma}) 
c^\dagger_{i,\overline{\sigma}}c^{\phantom{\dagger}}_{j,\overline{\sigma}} 
(1-n^{\phantom{\dagger}}_{j,\sigma})\nonumber \\ 
  &&\phantom{t_0\sum_{{ \langle i,j\rangle, \sigma}
      }\Big[}
  -n^{\phantom{\dagger}}_{i,\sigma} c^\dagger_{i,\overline{\sigma}}
c^{\phantom{\dagger}}_{j,\overline{\sigma}} n^{\phantom{\dagger}}_{j,\sigma}
+ \text{h.c.}
  \Big].
\end{eqnarray}

Likewise, hopping processes between third-nearest neighbor (3NN) sites, i.e., sites that
lie in-line on one of the axes and are separated by two links, are described by the 
contributions of
\begin{eqnarray}
\label{eq:3NN_hoppings}
  T''_0&=& t''\sum_{\mathclap{\langle \langle \langle i,j\rangle\rangle\rangle; 
\sigma}}  \Big[(1-n^{\phantom{\dagger}}_{i,\sigma}) 
c^\dagger_{i,\overline{\sigma}}c^{\phantom{\dagger}}_{j,\overline{\sigma}} 
(1-n^{\phantom{\dagger}}_{j,\sigma})\nonumber \\ 
  &&\phantom{t_0\sum_{{ \langle i,j\rangle, \sigma}
      }\Big[}
  -n^{\phantom{\dagger}}_{i,\sigma} c^\dagger_{i,\overline{\sigma}}
c^{\phantom{\dagger}}_{j,\overline{\sigma}} n^{\phantom{\dagger}}_{j,\sigma}
+ \text{h.c.}
  \Big].
\end{eqnarray}
In 1D, only the second type, i.e., the contribution $T_0''$, exists because there are no diagonals
so that the double-prime processes represent next-nearest neighbor hopping. 
Since this makes the nomenclature NN, NNN and 3NN  {ambiguous}
if 1D and 2D are considered both, we use the terms \emph{prime} and \emph{double-prime}
 hopping instead. In 2D, both exist; in 1D only the double-prime hopping.

Apart from these hopping processes further \emph{spin-dependent} hops
 can occur in the effective model. Whenever charges hop from one site to another, e.g., 
from $i$ to $j$, with a NN site $k$ in between, spin-dependent hops of the form
\begingroup
\allowdisplaybreaks
\begin{subequations}
\label{eq:spin_dependent_hopping}
\begin{alignat}{3}
	T'_{s,0}=t'_s &\sum_{\mathclap{\substack{\langle i,k,j\rangle\\ \alpha,\beta}}}
  \,\biggl\{
  \big[(1-n^{\phantom{\dagger}}_{i,\alpha}) c^\dagger_{i,\overline{\alpha}}
  \boldsymbol{\sigma}_{\overline{\alpha},\overline{\beta}}
  c^{\phantom{\dagger}}_{j,\overline{\beta}} 
  (1-n^{\phantom{\dagger}}_{j,\beta}) \big] \boldsymbol{\cdot} {\bf S}_k
  \nonumber \\ 
  &+\big[ n^{\phantom{\dagger}}_{i,\alpha} c^\dagger_{i,\overline{\alpha}} 
  \boldsymbol{\sigma}_{\overline{\alpha},\overline{\beta}}
  c^{\phantom{\dagger}}_{j,\overline{\beta}}
  n^{\phantom{\dagger}}_{j,\beta}\big] \boldsymbol{\cdot}
  {\bf S}_k +\text{h.c.} \biggr\} \\
  T''_{s,0}= t''_s &\sum_{\mathclap{\substack{\langle \langle i,k,j\rangle\rangle\\ \alpha,\beta}}}
  \,\,\biggl\{
  \big[(1-n^{\phantom{\dagger}}_{i,\alpha}) c^\dagger_{i,\overline{\alpha}}
  \boldsymbol{\sigma}_{\overline{\alpha},\overline{\beta}}
  c^{\phantom{\dagger}}_{j,\overline{\beta}} 
  (1-n^{\phantom{\dagger}}_{j,\beta}) \big] \boldsymbol{\cdot} {\bf S}_k
  \nonumber \\ 
  &+\big[ n^{\phantom{\dagger}}_{i,\alpha} c^\dagger_{i,\overline{\alpha}} 
  \boldsymbol{\sigma}_{\overline{\alpha},\overline{\beta}}
  c^{\phantom{\dagger}}_{j,\overline{\beta}}
  n^{\phantom{\dagger}}_{j,\beta}\big] \boldsymbol{\cdot}
  {\bf S}_k +\text{h.c.}  \biggr\}
\end{alignat}
\end{subequations}
\endgroup
 {occur}. 
 {These processes not only involve the hopping of a fermion over a nearest-neighbor, but also its 
interaction with the spin of this nearest-neighbor. For instance, the spin of the hopping fermion
may swap with the spin of the nearest-neighbor.}
Just like the hopping processes in  \eqref{eq:NNN_hoppings} and \eqref{eq:3NN_hoppings},
 the spin-dependent processes \eqref{eq:spin_dependent_hopping} do not change the overall number of DOs.
As before, in 1D only the double-prime processes exist because of the lack of diagonals.
The leading orders of the contributions that emerge in this process may be determined 
analytically via perturbation-theoretical approaches \cite{mielk91aa,eskes94b} or 
numerically by means of the above-discussed CUT \cite{Reischl2004}.

In the following, we make use of the values derived analytically for 1D in \refcite{mielk91aa} which read
\begin{subequations}
\begin{alignat}{4}
  t''&=-\frac{t_0^2}{2U} \\
  t_s''&=\frac{t_0^2}{U}.
\end{alignat}
\end{subequations}
 {A generalization of these contributions to the additional processes arising in 2D is easily possible.
In 2D, there is exactly one shortest route from $i$ to $j$ which can generate a $t''$ contribution.
For diagonal hopping, i.e., for $t'$, there are two shortest routes. A diagonal step on a square lattice can happen via first 
a horizontal step and then a vertical step or vice-versa. 
For spin-independent diagonal hopping, both routes contribute and hence we have a factor 2}
\begin{equation}
  \label{eq:generalisation_tp}
  t'=-\frac{t_0^2}{U}.
\end{equation}
 {For spin-dependent processes the involved  intermediate lattice site $k$ distinguishes
the two routes so that no doubling is needed}
\begin{equation}
  \label{eq:generalisation_tsp}
  t_s'=\frac{t_0^2}{U}.
\end{equation}
 {These leading contributions \eqref{eq:generalisation_tp} and \eqref{eq:generalisation_tsp} generalized for 2D 
are consistent with the numerically determined contributions $t'$ and $t_s'$ of comparable studies, cf.\ \refcite{Reischl2004}.
They also agree with the 2D results in \refcite{eskes94b}.} 

 {So far only the charge degrees of freedom are considered with $H_\mathrm{0,eff}$. Next, we extend the model by
the spin-spin interactions. This model is equivalent to what is called \tjm{} in the literature, except that
we include the interaction term $H_\text{int}$ to keep track of the energy shifts.
The additional Heisenberg contribution reads}
\begin{equation}
  H_J = J\sum_{\mathclap{\langle i,j\rangle}} \vec{S}_i \vec{S}_j = 
	\frac{J}{2} \sum_{\mathclap{\langle i,j\rangle}} \left(P_{ij}-\frac{1}{2}\right)
\end{equation}
with $J=\nicefrac{4t_0^2}{U}$  {so that the  effective Hamiltonian becomes }
\begin{equation}
  \label{eq:full_tjm}
  H_\mathrm{eff}=H_{0,\mathrm{eff}}+H_J.   
\end{equation}
 {The alternative notation employing the permutation operator $P_{ij}$, which interchanges two spins on the 
lattice sites $i$ and $j$ is equivalent to the spin-spin exchange
for $S=\nicefrac{1}{2}$. It is particularly useful in numerics where quantum mechanical states are
represented by bit patterns.}

 {Recalling the bandwidth $W=2zt_0$ with the coordination number $z$ of the lattice 
helps us to identify the physically relevant parameter choices.
Inserting it into the leading contributions above, it is easy to identify relevant physical regimes. 
We focus on two paramter sets (A) and (B).
The first is motivated from the application to cuprates 
where $J=\nicefrac{t_0}{3}$ is a representative value. 
The second parameter set (B) is theoretically motivated. It represents roughly the boundary value
$U=W$ up to which the mapping from the Fermi-Hubbard model to the \tjm{} is reasonable
\cite{Reischl2004,hamer10}. For lower values of the interaction the assumption of split
Hubbard bands is no longer justified. Case (B) is of particular interest because it represents
the limiting case with maximum second-order terms in the \tjm. }
For ease of identification, we use the abbreviations (A) and (B) below to distinguish 
 {between these parameter sets. We emphasize that in case (B)  different second order terms
occur} depending on the dimension $d$ of the system.  {For instance, for the magnetic coupling we have}
\begin{subequations}
\label{eq:parameter_regimes}
\begin{alignat}{3}
  (A)&~~~ J=\frac{t_0}{3} \label{eq:parameter_regimes_A}\\
  (B)&~~~ J=\frac{t_0}{d}. \label{eq:parameter_regimes_B}
\end{alignat}
\end{subequations}

\section{Methods} 
\label{s:method}

In this section, we present a brief overview over the methods used to calculate the quantities 
 {computed} in this article. Importantly, we point out the 
strengths and shortcomings of the techniques used. 
In-depth  derivations can be found in the references given.

 {The lower band edge of the Hubbard band of the hole is a key quantity.}  {If it
falls below $U/2$ the assumed Mott insulator is instable. 
The determination of $\omega_\text{min}$} is achieved using two fundamentally different approaches. 
In the first approach using iterated equations of motion (\ieom{}), we obtain the energy spectrum of the 
system and consequently have direct access to the minimum energy $\omega_\mathrm{min}$ of the 
lowest-lying excitation. In the second approach, we simulate the full dynamics of the hole-doped \tjm{} 
in a numerically exact manner by means of  {the} Chebyshev expansion technique (\cet{}) providing the 
spectral function of the initial hole excitation.

The two methods appear similar at first glance, but the crucial difference resides in the fact that 
the \ieom{} approach works in the Heisenberg picture addressing operators while
the Chebyshev expansion treats quantum states. The \ieom{} approach 
systematically truncates the underlying Hilbert space of operator monomials, but 
treats  {the thermodynamic limit of an infinite lattice}.
The \cet{} considers the whole  Hilbert space with exponentially increasing dimension  {for increasing system size}.
This requires to consider finite systems.   
For the \ieom{} approach, no simulation of the time dependence of the hole-doped 
\tjm{} up to a specific threshold time $t_\mathrm{max}$ needs to be performed. Instead, 
the excitation spectrum can be deduced  {directly} by diagonalization. 
We start by a dedicated analysis of \ieom{} in \Cref{ss:ieom} 
before the numerical approach of \cet{}  is presented in \Cref{ss:cet}.

\subsection{Iterated equations of motion}
\label{ss:ieom}

In order to deduce the full energy spectrum  {and the lower band edge in particular}
 we resort to the iterated equations of
 motion approach \cite{Uhrig2009,Hamerla2013,Hamerla2014}, 
a brief summary of which is given in the first part of this section. 
The second part is dedicated to the necessary modifications of the method which warrant 
a unitary time evolution on the operator level \cite{Kalthoff2017,Bleicker2018}, 
and the third part describes the concrete application of the \ieom{} to the \tjm{}.

We start by considering an arbitrary operator in the Heisenberg picture
\begin{equation}
	\label{eq:op_expansion}
	A(t)=\sum_i h_i(t)A_i.
\end{equation}
Here, all time dependence is contained in the complex prefactors $h_i(t)$; the 
constant operators $A_i$ from the Schr\"odinger picture form an operator basis. In the following, 
$\hbar$ is set to unity for simplicity. The linear independence of the $A_i$ is required
since otherwise the above expansion \eqref{eq:op_expansion} would not be unique.
Without explicit time dependence of the Hamiltonian 
the Heisenberg equation of motion becomes
\begin{equation}
	\label{eq:h_eom}
	\dv{t} A(t) = i\comm{H(t)}{A(t)} =: i \Liouville{A(t)}
\end{equation}
with the Liouville superoperator $\Liouville{\cdot}$. 
Then, inserting \Cref{eq:op_expansion} into \eqref{eq:h_eom} leads to
\bes
\begin{align}
	\label{eq:eom_op_expansion}
	\dv{t}A(t) &=i\Liouville{A(t)}\\
	& = i\sum_i h_i(t)\Liouville{A_i}.
\end{align}
\ees
It is possible to expand all operators $\Liouville{A_i}$ in the chosen basis $\{ A_i\}$ by
\begin{equation}
	\label{eq:liouville_auf_op_expansion}
	\Liouville{A_i} := \sum_j M_{ji} A_j
\end{equation}
leading to the Liouvillian matrix $\matrixsymbol{M}$, also called dynamic matrix.  
For a compact notation, it is advisable to
combine the time dependent prefactors $h_i(t)$ to a vector $\vec{h}(t)$.
Its dynamics is given by
\begin{equation}
\label{eq:matrix_dgl}
	\dv{t}\vec{h}(t) = i \matrixsymbol{M} \vec{h}(t).
\end{equation}

For the computation of the Liouvillian matrix, it is  {convenient} to use an orthonormal 
operator basis $\{A_i\}$ (ONOB) so that each matrix element can be computed directly by
\begin{equation}
	\label{eq:matrix_elements_onb}
	M_{ji}=\scalarpr{A_j}{\Liouville{A_i}}.
\end{equation}
It has been previously shown \cite{Kalthoff2017,Bleicker2018} that it is crucial to achieve Hermiticity of 
$\matrixsymbol{M}$, i.e., preserving the property $M_{ji} = M_{ij}^*$ in order to obtain
oscillatory solutions. Otherwise, exponentially increasing solutions are possible and will occur.
They definitely do not reflect physical behavior.
The Hermiticity of $\matrixsymbol{M}$ is tantamount to $\Liouville{\cdot}$ being self-adjoint. 
To achieve this property, one has to use a suitable
operator scalar product for two linear operators $A$ and $B$ 
defined on a locally finite-dimensional Hilbert space $\mathcal{H}$, i.e., 
$\operatorname{dim}(\mathcal{H}) < \infty$. An  {advantageous} choice is the Frobenius scalar product
\begin{equation}
	\label{eq:frobenius_scalar_product}
	\scalarpr{A}{B} := \normfactor \Tr(A^\dagger B)~\text{with}~\normfactor
	:=\frac{1}{\Tr(\mathbb{1})}.
\end{equation}
The prerequisite of a locally finite-dimensional Hilbert space clearly holds for all spin systems and  
all fermionic systems such as the Fermi-Hubbard model  {or models like the \tjm{}} with both spin 
and fermionic degrees of freedom. Bosonic degrees of freedom are excluded due to their locally 
infinite-dimensional Hilbert spaces.

Note that the scalar product \eqref{eq:frobenius_scalar_product} can also be interpreted
physically since it equals 
the high-temperature limit $T\to\infty$ of the thermal expectation value
\bes
	\label{eq:high_temperature_limit_frobenius}
\begin{align}
	\scalarpr{A}{B} &= \lim_{T\to\infty} \expval{A^\dagger B}
	\\
	&= \lim_{T\to\infty} \Tr \left(\rho A^\dagger B\right)
\end{align}
\ees
in the canonical ensemble for a density matrix $\rho = \nicefrac{e^{-\beta H}}{Z}$ 
with the partition sum $Z=\Tr\left(e^{-\beta H}\right)$ and the 
inverse temperature $\beta \ge 0$. Using this very scalar product ensures that 
$\Liouville{\cdot}$ is indeed self-adjoint and thus that the dynamic matrix is Hermitian. 
This stems from the invariance of the trace under cyclic permutations in \eqref{eq:matrix_elements_onb}, 
see \refcite{Kalthoff2017} and especially \refcite{Bleicker2018}.

Next, one has to choose an appropriate operator basis $\{A_i\}$ to describe the dynamics of a hole 
excitation in the half-filled \tjm{}. Generally, there are various techniques to do so by either 
resorting to the iterative approach of looping operators, cf., especially 
\refcite{Hamerla2013,Hamerla2014}, or by using a closed
operator basis which has to be constructed \textit{a priori} as was done in, e.g., 
\refcite{Bleicker2018}. The advantage of the first approach is that it considers more operators 
relevant to the actual dynamics while the advantage of the second approach is that it is simpler 
to ensure the orthonormality of the operator basis  {since it is constructed} beforehand. 

In the following, we present a  {mixed} approach that combines the strengths of both techniques. 
We recall that for two bounded operators  $A_1$ and $A_2$ acting on two different  Hilbert spaces 
$\mathcal{H}_{\!1}$ and $\mathcal{H}_{\!2}$ the  trace in the product space 
$\mathcal{H} = \mathcal{H}_{\!1} \otimes \mathcal{H}_{\!2}$ can be split  {into two factors}
\begin{equation}
     \label{eq:hilbert_space_split}
     \Tr\left(A_1 A_2\right) = \Tr_1\left(A_1\right) \Tr_2\left(A_2\right),
\end{equation}
where $\Tr_m(.)$ denotes the partial trace over $\mathcal{H}_{\!m}$. 
Hence, in the Hilbert space of an $N$-site lattice 
the trace of a product of operators acting on different sites can be factorized 
into a product of local traces in the four-dimensional local Hilbert space spanned by
 $\left\{ \ket{0},\ket{\uparrow},\ket{\downarrow},\ket{\uparrow\downarrow}\right\}$. 
This fact helps us to establish an unambiguous representation of operators. 
Each  {given} operator can be decomposed into a product of operators 
acting on different sites where the 16 local operators 
listed in \Cref{tab:operator_basis} form a local orthogonal basis.
The operator $\nU=(n_\downarrow-1/2)(n_\uparrow-1/2)$
 measures the deviation from half-filling at a given site:
at half-filling $\nU=-1/4$ holds, in presence of a DO (electron or hole) 
$\nU=1/4$ holds. Note that  {here} site indices and the resulting normalization factors 
are omitted for brevity. We stress that 
the local operators in \Cref{tab:operator_basis} are mutually orthogonal, but not normalized.
\begin{table}[!h]
	%\vspace{.5cm}
    \centering
    \begin{tabular}{| c | c | c | c |}
    %&    $\ket{0}$      &$\ket{\up}$            & $\ket{\down}$      & $\ket{\up\down}$\\
                    \hline 
         $\phantom{\Big|}\mathbb{1}$ & $\nU$ &  $\splus =\splus*$ & $\cc$ \\[1mm] \hline
         $\phantom{\Big|}n_\downarrow -1/2$  & $n_\uparrow -1/2 $ & $\sminus =\sminus*$ & $\ctct$ \\[1mm] \hline
         $\phantom{\Bigg|}\dtdown=\dtdown*$ & $\dtup=\dtup*$ & $\ltdown=\ltdown*$ & $\ltup=\ltup*$ \\[1mm] \hline
         $\phantom{\Bigg|}\ddown=\ddown* $ & $\dup=\dup* $ & $\ldown=\ldown*$ & $\lup\!=\!\lup*$ \\[1mm] \hline
        \end{tabular}  
    \caption{Operator basis for a four-dimensional local Hilbert space 
		being orthogonal with respect to \eqref{eq:frobenius_scalar_product}. 
		The overbars indicate that the creation operators imply a certain projection 
		compared to standard fermionic creation operators.
		As denoted, the operators are not normalized.}
    \label{tab:operator_basis}
    %\vspace{-.3cm}
\end{table}

In order to describe the dynamics of a hole inserted into the disordered spin background we consider the time evolution of the operator $\herz{i\uparrow}(t)$. Initially, $\herz{i\uparrow}(t\!=\!0) = \herz{i\uparrow}$ holds
so that the initial condition for the prefactors $h_i$ 
in \eqref{eq:op_expansion}  reads
\begin{equation}
 \label{eq:h_initial_condition}
 h_i(0) =
  \begin{cases} 
      1 \hfill & \text{ if $i=1$} \\
      0 \hfill & \text{ otherwise,}
  \end{cases}
\end{equation}
 {setting}  $A_1:=\herz{i\uparrow}$.
Starting from $\herz{i\uparrow}$ the operator basis $\{A_i\}$ is constructed by repeatedly applying the 
Liouville superoperator to the current basis operators, simplifying the results so that the local 
operators are one of  \num{16} local operators in \Cref{tab:operator_basis}. In this way, 
an operator of the basis is constructed as operator monomial, i.e., a product of the local operators
on a certain subset of sites of the lattice.
The monomials which are created for the first time by the current iteration extend 
the basis. The total number of applications of $\Liouville{\cdot}$ to the basis operators 
is called the order $m$  {of the iterative extension of the
basis of monomials. We call one iteration of $\Liouville{\cdot}$ a \emph{loop}. 
This means, for instance, that three commutations of the Hamiltonian $H$ 
 yield the so called $3$-loop basis. Generally, $m$ iterations of $\Liouville{\cdot}$
lead to the $m$-loop basis.}

\begin{table}[!h]
	%\vspace{.5cm}
    \centering
    \begin{tabular}{| c | c | c | c |}
    %&    $\ket{0}$      &$\ket{\up}$            & $\ket{\down}$      & $\ket{\up\down}$\\
                    \hline
         $\phantom{\Big|}\mathbb{1}$ & $\sz=\sz*$   & $\splus $ & $\sminus$  \\[1mm] \hline
         $\phantom{\Big|}\dtdown $ & $\ltdown$  & $\dtup $ & $\ltup$ \\[1mm]
         \hline
        \end{tabular}  
    \caption{Reduced operator basis for  {the} two-dimensional local Hilbert space. 
		All  {pairs of} operators are orthogonal with respect to \eqref{eq:frobenius_scalar_product}, 
		but  {the operators} are not normalized.}
    \label{tab:2d_operator_basis}
    %\vspace{-.3cm}
\end{table}

Our aim is to capture the dynamics of a single hole, i.e., a DO of holes, or a single DO of 
particles at half-filling. 
The above procedure, however, generates operators which are effective  {on} an increasing number of
DOs. The corresponding monomials are relevant if  {larger levels of doping} are considered. But for
the particular goal here they are detrimental in two respects: (i) they represent a
computational burden leading to unnecessarily large operator bases; 
(ii) they correspond to processes which
cannot take place at half-filling leading to spurious eigenvalues of the dynamic matrix
$\matrixsymbol{M}$.  {The corresponding eigenvectors do not
matter at half-filling. Hence, it is indicated to discard the monomials which
are effective only for two or more DOs and we restrict the tracked operator monomials to
those creating a single hole or a single DO}. This means that  {in the relevant monomials
there is only} a single site with a hole/DO creation operator 
$\overline{h}^{\dagger}_\sigma$ or $\overline{d}^{\dagger}_\sigma$. 
At all other sites, where the operator monomial has a non-trivial effect, only
operators which conserve the number of DOs \emph{and} which have a non-zero
effect at half-filling appear. These are the operators of the upper row in \Cref{tab:2d_operator_basis}.
The lower row in this table lists the operators which create the hole/DO at the 
one site of the operator monomial.  {Charge hopping} processes can only occur 
at this particular site. It is sufficient that the normalization factor is calculated 
by the trace in the reduced half-filled Hilbert space, i.e., summing over the two local states 
$\ket{\uparrow}$ and $\ket{\downarrow}$ only.

For the  {iterative} construction of the operator basis
we apply the Liouvillean, corresponding to the commutation with $H_\text{eff}$,
with or without the prime and double-prime terms recursively. The resulting operators are expanded
in operator monomials of which only those are kept which create a DO at one site
and elsewhere only consist of operators of the upper row in \Cref{tab:2d_operator_basis}. This
reduces the computational effort considerably and provides the physically relevant dynamic
matrix $\matrixsymbol{M}$. The identity operator $\mathbb{1}$ does not need to be tracked.
Once the operator basis of monomials  {is determined, the dynamic
matrix $\matrixsymbol{M}$ is calculated using \eqref{eq:matrix_elements_onb} and 
diagonalized leading to the desired set of eigenvalues $\omega_n$ and 
the corresponding eigenvectors $\vec{v}_n$}.

Given the enormous size of the underlying Hilbert space  {of operators, i.e., the enormous size of} the dimension of $\matrixsymbol{M}$, it is necessary to resort to efficient diagonalization techniques such as the 
Arnoldi iteration \cite{Arnoldi1951} which simplifies to the well-known 
Lanczos algorithm \cite{Lanczos1950} in case of Hermitian matrices. 
Since the Hermiticity of $\matrixsymbol{M}$ is guaranteed by construction 
the Lanczos algorithm can be employed to compute the (reduced) $f$-dimensional Krylov space and 
the corresponding set of eigenvalues and eigenvectors. We denote these reduced sets by $\overline{\omega}_n$ and 
$\overline{\vec{v}}_n$, respectively. With the help of these sets  {the dynamics of the system can be expressed 
to very good accuracy} as a linear combination according to
\begin{equation}
  \label{eq:matrix_dgl_solution}
  \vec{h}(t)=\sum_{n=1}^f \alpha_n e^{i\overline{\omega}_n t}\overline{\vec{v}}_n
\end{equation} 
with the coefficient set $\alpha_n$ chosen in such a way that the initial condition given by \Cref{eq:h_initial_condition} is fulfilled.  {We varied the dimension $f$ of the Krylov space to monitor if any 
changes in the results occur. We finally chose $f=200$ 
from which on no changes can be discerned anymore. In particular, the minimum of $\overline{\omega}_n$
represents a very reliable estimate for the band edge. As long as $U/2+\overline{\omega}_n >0$, the Mott insulating phase
is locally stable.}

Since we consider a fully disordered spin background
 we do not deal with a pure state but with  {a mixed ensemble corresponding to} 
the high-temperature limit of the canonical ensemble $\rho\propto e^{-\beta H}$, i.e., 
to the density matrix
\begin{equation}
  \label{eq:density_matrix}
  \rho_0 \propto \mathbb{1}. 
\end{equation}
We insert a hole into it at time $t\!=\!0$ using the respective creation operator 
$\overline{h}^{\dagger}_{i\uparrow}$. 
The dynamics of this charge excitation is described by the retarded Green's function
\begin{equation}
	\label{eq:raw_autocorrelation_function}
  g(t) = -i \Tr\left(\overline{h}_{i\uparrow}(t)
	\overline{h}^{\dagger}_{i\uparrow}\rho_0\right)\theta(t)
\end{equation}
where $\theta(t)$ stands for the Heaviside function. No commutator appears because
the corresponding term vanishes  {since} no hole can be annihilated 
in the exactly half-filled state.

 {In the framework of the iEoMs the above retarded Green's function is found by} inserting the operator expansion 
\eqref{eq:op_expansion} into \eqref{eq:raw_autocorrelation_function} twice, 
once for the creation and once for the annihilation operator  {leading to}
\begin{subequations}
\label{eq:autocorrelation_ieom}
\begin{alignat}{3}
  g(t)  &= -i \expval{\overline{h}_{i\uparrow}(t)\overline{h}^{\dagger}_{i\uparrow}}\theta(t) 
	\\
  		&= -i\sum_{mn} h_m(t) h_n^*(0) \expval{A_m A_n^\dagger}\theta(t) 
			\\
  		&= -i\sum_n \left|\alpha_n \right|^2 e^{i \overline{\omega}_n t}\theta(t) 
			\label{eq:autocorrelation_ieom_2}.
\end{alignat}
\end{subequations}
Note that $\expval{A_m A_n^\dagger}=\delta_{mn}$ holds since the corresponding operator basis 
$A_m$ is orthonormal with respect to \eqref{eq:frobenius_scalar_product}.
The spectral density $A(\omega)$ can be obtained from the Fourier transform $g(\omega)$ of 
\eqref{eq:autocorrelation_ieom_2}
\begin{subequations}
\begin{alignat}{3}
  \label{eq:ieom_spectral_density}
  A(\omega) &= -\frac{1}{\pi} \Im g(\omega) \\
  			&= \sum_n \abs{\alpha_n}^2 \delta\left(\omega-\overline{\omega}_n\right).
\end{alignat}
\end{subequations}
 {As expected, the modulus squared of the coefficients $\alpha_n$ indicates the weight and hence the relative
importance of the corresponding process for the hole dynamics.}

 {A finite number of iterations, i.e., a finite order $m$, implies that only processes 
with a finite spatial spread in the infinite system are taken into account.
This implies that only a finite number of eigenvalues $\omega_n$ occurs so that the spectral density 
\eqref{eq:ieom_spectral_density} is not continuous, but consists of discrete $\delta$-spikes.
This discreteness also results from the use of a finite-dimensional Krylov space in the Lanczos
diagonalization. But this effect is straightforward to control because the Krylov dimension $f$
can easily be increased by for instance, a factor 2. The finiteness, however, of the loop order $m$
cannot be increased easily. In order to plot spectral densities and to compare them
between different approaches, we broaden the $\delta$-spikes artificially 
by Gaussians according to}
\begin{equation}
  \label{eq:ieom_spectral_density_full}
  A(\omega) = \sum_n \frac{\abs{\alpha_n}^2}{\sqrt{2\pi}\sigma}
	\exp\left(-\frac{(\omega-\overline{\omega}_n)^2}{{2\sigma^2}}\right).
\end{equation}
 {The artificial broadening $\sigma$ has the unit of an energy, recall $\hbar=1$, and 
its value will be discussed below.}

\subsection{Chebyshev expansion technique}
\label{ss:cet}

As many numerical approaches, the \cet{} needs a finite-dimensional Hilbert space
so that only finite clusters can be  {dealt with}. This calls for a suitably chosen complete set 
of orthonormal states $\left\{\ket{i}\right\}$ forming a basis of the Hilbert space $\mathcal{H}$ 
for the Hamiltonian \eqref{eq:generalized_tj_model}.  {In contrast to the previous section, we
stress that CET works in the Schr\"odinger picture so that the states are the usual kets.}
To increase the overall performance we resort to an integer representations of the basis states.
We exemplify this procedure in the following before discussing how to derive Green's functions from the 
Hamiltonian matrix. Consider a \tjm{} doped with a single hole where exactly one hole is inserted 
into a lattice of $N$ sites. Then, the dimension of the Hilbert space is 
$\dim\left(\mathcal{H}\right)=N\,2^{N-1}$ since the hole may occupy one of $N$ sites while on all 
remaining sites the spins $\nicefrac{1}{2}$ can point  {upwards} or downwards.  {For notational
simplicity, we} artificially enlarge the basis size to $\Dim=N\,2^N$ states while keeping in mind 
that the spin orientation at the site occupied by the hole  {has no physical meaning}.
In this way, a real space basis of the binary form 
\begin{equation}
	\ket{i}=\ket{i_{N-1}\ldots i_0}\ket{h_{N-1}\ldots h_0}	
\end{equation}
can be constructed. Here, all spins can be either orientated upwards, i.e., 
$\uparrow\,\equiv\!0$, or downwards, i.e., $\downarrow\,\equiv\!1$, so that the relation 
$i_j\in\left\{0,1\right\}$ holds. The hole always occupies exactly one site $h_k=1$ with all 
remaining sites being empty such that $h_j=0\,\forall\,j\neq k$. This allows for an easy and 
concise identification of a given basis state $\ket{i}$ by the integer representation 
$I\in{0,\ldots,N\,2^N - 1}$ using
\begin{equation}
	I = k\,2^N + \sum_{j=0}^{N-1} i_j 2^j
\end{equation}
where the last sum is equal to the integer value of the binary number given 
by the binary pattern of the spin orientations.
As long as we do not consider the magnetic exchange  {$H_J$} the dynamics in the \tjm{} \emph{only}
occurs at the  {hole position} $k$. This facilitates the numerical task considerably. It is possible to 
construct the respective Hamiltonian matrix either on-the-fly by algorithms linear in the basis size or
 to keep a highly sparse copy of it.

In order to compute spectral densities with the CET, we determine the retarded Green's function $g(t)$ 
in a first step. The trace over the half-filled Hilbert space in 
\eqref{eq:raw_autocorrelation_function} 
has to be taken which could strongly limit  this approach. 
Fortunately, stochastic trace evaluation as initially proposed by 
Skilling \cite{Skilling1988} and later generalized by others 
\cite{Drabold1993,Silver1994,weiss06a} can be employed here. 
It consists of approximating the full trace 
$\Tr(A)$ by $R\ll \Dim$ randomly chosen quantum states. 
Using a set of $R$ normalized states $\ket{r}$ whose complex coefficients are each 
drawn from a normal distribution we approximate traces by the average of the
 expectation values
\begin{equation}
	\label{eq:trace_estimate}
	\Tr\left(O\right) = \Dim \overline{\mel{r}{O}{r}}
\end{equation}
where the overbar denotes the process of determining the arithmetic average over
the $R$ random states $\left\{\ket{r}\right\}$ and $\Dim$ 
is the dimension of the half-filled Hilbert space. 
The standard deviation of the estimate \eqref{eq:trace_estimate} 
scales like $1/\sqrt{R\Dim}$. Finally, inserting \eqref{eq:trace_estimate} and \eqref{eq:density_matrix} 
into \eqref{eq:raw_autocorrelation_function} leads to the approximated retarded Green's function
\begin{equation}
  \label{eq:autocorrelation_function}
  g(t) \approx -\frac{i}{R} \sum_{r=1}^R \mel{r}{e^{iH_\text{eff}t}
  \overline{h}_{i\uparrow}e^{-iH_\text{eff}t}\overline{h}^{\dagger}_{i\uparrow}}{r} \theta(t).
\end{equation}
 {Below, we consider the hole dynamics in both the complete \tjm{} in its form $H_\mathrm{eff}$ as well as the 
hopping-only model $H_{0,\mathrm{eff}}$. The last case, i.e., setting the magnetic exchange to zero with $J=0$, allows 
for a significant simplification of \eqref{eq:autocorrelation_function} implying}
\begin{equation}
  \label{eq:autocorrelation_function_without_J}
  g_{0,\mathrm{eff}}(t) \approx -\frac{i}{R} \sum_{r=1}^R \mel{r}
      {\overline{h}_{i\uparrow}e^{-iH_{0,\text{eff}}t}\overline{h}^{\dagger}_{i\uparrow}}{r} \theta(t),
\end{equation}
where we exploit that $e^{iH_\text{eff}t}$ for $J=0$ has no impact on 
$\bra{r}$ because no hopping can take place 
in $\bra{r}$ such that this contribution in \eqref{eq:autocorrelation_function}  can be omitted in 
\eqref{eq:autocorrelation_function_without_J}. 

The Green's functions \eqref{eq:autocorrelation_function} and \eqref{eq:autocorrelation_function_without_J} are 
calculated for a finite time span $\left[0;t_\mathrm{max}\right]$ in time steps of $\dd{t}$. Thereafter, they 
are Fourier transformed 
\begin{equation}
  \label{eq:manual_fft}
  g(\omega):=\sum_n e^{-i\omega t_n} g(t_n) \dd{t}.
\end{equation}
 {For the sake of notational brevity, we  use  the same symbol $g$ for time and for frequency dependence.} 

The finite time interval leads to spurious phenomena in the Fourier transforms \eqref{eq:manual_fft} which can be 
systematically suppressed by damping the temporal Green's function $g(t)$  {by multiplying it with
a decreasing function. For simplicity, we opt for the approach 
to damp the Green's function by means of 
\begin{equation}
  \label{eq:gauss_time_domain}
  \widetilde{g}(t) = g(t)\exp\left(-\nicefrac{1}{2}\cdot\sigma^2 t^2\right).
\end{equation}
We recall that the multiplication in the time domain \eqref{eq:gauss_time_domain} is equal to the convolution of 
$g(\omega)$ in the frequency domain with the Gaussian kernel 
$K\propto\exp\left(\nicefrac{-\omega^2}{2\sigma^2}\right)$. }
Finally, the spectral density of the single hole excitation is obtained by
\begin{equation}
	\label{eq:spectral_function}
	A(\omega) = -\frac{1}{\pi}\Im \widetilde{g}(\omega).
\end{equation}

To compute the Green's function, we need the time-dependence $\ket{\psi(t)}$ 
of the randomly chosen  {initial} state after inserting the hole 
\begin{equation}
	\ket{\psi(t)}:=\exp\left(-iHt\right) \overline{h}^\dag_{i\uparrow}\ket{r}	.
\end{equation}
 {To do so} we resort to the Chebyshev expansion 
technique \cite{Tal-EzerH.1984} which consists of the expansion of the unitary 
time evolution operator $U(t)=\exp(-iH_\text{eff}t)$ in terms of Chebyshev polynomials
\begin{subequations}
\label{eq:chebyshev_polynomials_recursion_relation}
\begin{alignat}{3}
  T_0(y) &= 1, \qquad T_1(y)=y \\
  T_{n+1}(y) &= 2y T_n(y)-T_{n-1}(y)
\end{alignat}
\end{subequations}
which are defined on the closed interval $I=\left[-1;1\right]$. To be able to apply this
 technique to a general Hamiltonian $H$ a finite rescaling 
$H\to H' = (H-b)/a$ is necessary to ensure that the spectrum of $H'$ lies
in the interval $I$.  For this rescaling one needs
 an estimate  of the extremal eigenvalues \cite{Lanczos1950,Arnoldi1951,Kuczynski1992} 
of $H$ to obtain $a=\nicefrac{1}{2}\left(E_\mathrm{max}-E_\mathrm{min}\right)$ 
and $b=\nicefrac{1}{2}\left(E_\mathrm{max}+E_\mathrm{min}\right)$.
Note that estimates in the form of upper bounds for $E_\mathrm{max}$ and lower bounds
for $E_\mathrm{min}$ are sufficient to warrant that the spectrum lies in $I$. 
Finally, the time-evolution operator  becomes
\begin{subequations}
\label{eq:cet_time_dependent_series_final}
\begin{alignat}{3}
  U(t) &= \sum_{n=0}^\infty \alpha_n(t)T_n(H') 
  \\
  \alpha_n(t) &= (2-\delta_{n,0}) i^n e^{-i b t} J_n(at) 
  \label{eq:cet_time_dependent_series_final_coefficients}
\end{alignat}
\end{subequations}
where the time-dependence is  {embodied in the}
Bessel functions of the first kind $J_n(at)$.
Eventually, the dynamics of an initial state $\ket{\psi_0}$ is given by
\begin{equation}
  \label{eq:cet_time_evolution_state}
  \ket{\psi(t)}=U(t)\ket{\psi_0}=\sum_{n=0}^\infty \alpha_n(t)
  \underbrace{T_n(H')\ket{\psi_0}}_{=:\,\ket{\phi_n}}
\end{equation}
with the basis states of the expansion $\ket{\phi_0}\!=\!\ket{\psi_0}$ 
and $\ket{\phi_1}=H'\ket{\psi_0}$ as well as $\ket{\phi_{n+1}}=
2H'\ket{\phi_n}-\ket{\phi_{n-1}}$.

Numerically, the infinite series is cut off at some finite, but large value 
$N_\mathrm{c}<\infty$. The time dependence of the prefactors 
 {resides in the Bessel functions $J_n(t)$ \cite{Olver2019}. } 
The higher the order $n$ the longer the Bessel function $J_n(t)$ takes to 
contribute noticeably to the series. Hence, an estimate for the
accuracy of the truncated series with cut-off $N_\mathrm{c}$ is given by
\begin{equation}
  \epsilon \lessapprox \left(\frac{a t_\text{max} \cdot e }{2 N_\mathrm{c}}\right)^{N_\mathrm{c}}. 
\end{equation}
Consequently, the truncation error is not only related to $N_\mathrm{c}$, but depends
also  on the maximum time up to which results are calculated as well as on the 
parameter $a$ which equals half the width of the energy spectrum.
An advantageous feature of the Chebyshev expansion is that 
increasing $N_\text{c}$ linearly increases the time $t_\text{max}$ 
up to which the error estimate remains the same.

\section{Real time dependence and band edges}
\label{s:method_comparison}

\subsection{Method comparison}

In order to understand the two methods described in \Cref{s:method} better and to 
see their strengths and weaknesses we apply them to the full \tjm{} as given by \Cref{eq:full_tjm}
on the 1D chain with the parameters (A) in \Cref{eq:parameter_regimes_A}.
In all calculations, the hopping element $t_0$ defines the energy unit and the time unit according to 
$[t]=\nicefrac{1}{t_0}$. 

\begin{figure}[ht]
  \centering
  % left bottom right top
  \includegraphics[trim=10 15 45 45,clip,width=.5\textwidth]{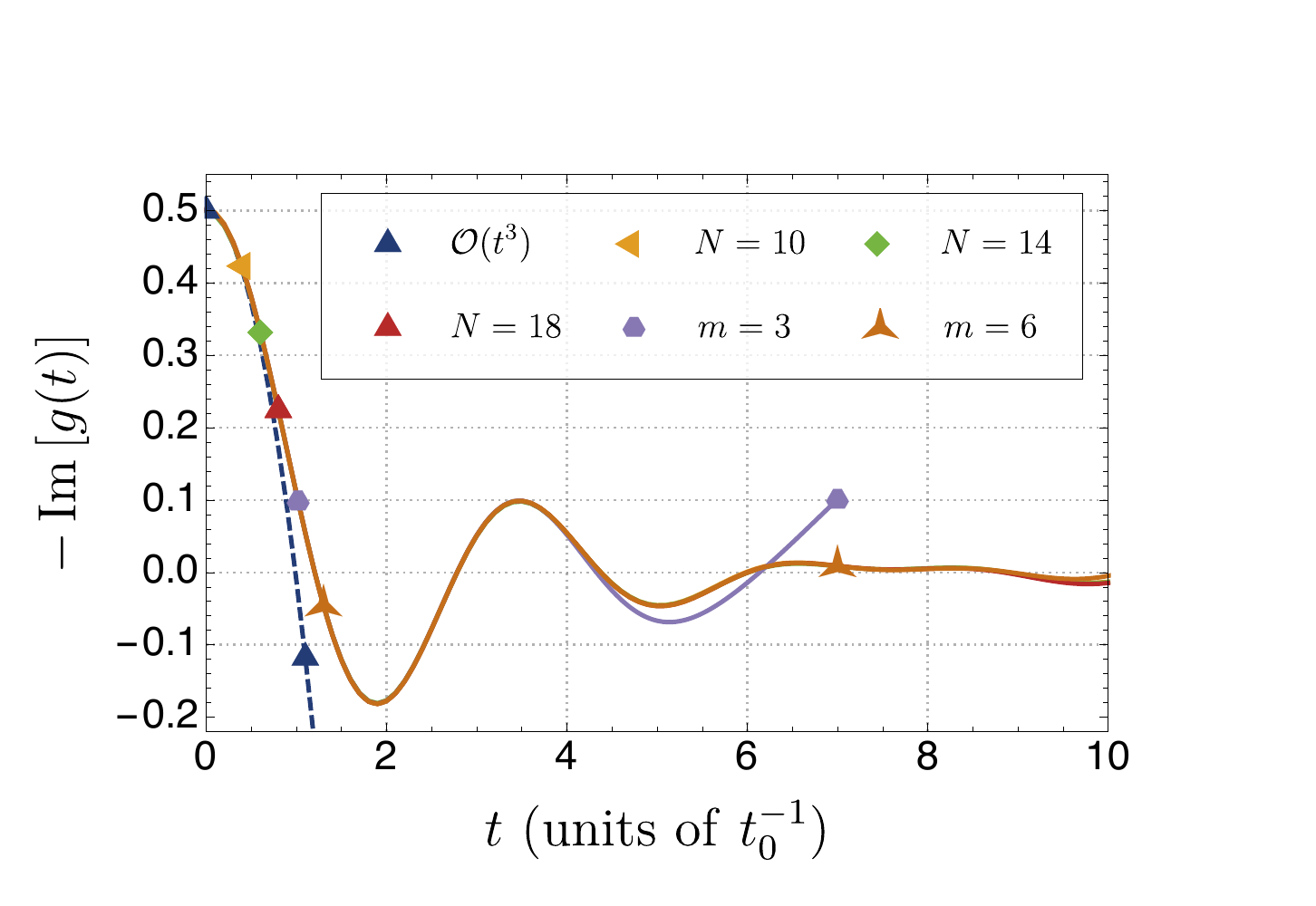}
  \caption{Retarded Green's function of the 1D chain for the full \tjm{} with parameters (A) as given in 
	\Cref{eq:parameter_regimes_A}. Results are determined by \eqref{eq:autocorrelation_ieom} and 
	\eqref{eq:autocorrelation_function} for various chain lengths $N$ (CET) and loop orders $m$ (iEoM). 
	The parabola of the analytical expansion, cf. \Cref{app:taylor}, in powers of $t$ up to $\mathcal{O}(t^3)$ 
	is depicted as well (blue triangles). Note that the iEoM result for $m=6$ loops first starts to deviate from the CET results at about $t\geq9/t_0$.}
  \label{img:paper2020_1D_A_comparison_CET_IEOM}
\end{figure}

 {Using \eqref{eq:autocorrelation_ieom} and \eqref{eq:autocorrelation_function}, 
the results for the retarded Green's function $g(t)$ 
are presented in \Cref{img:paper2020_1D_A_comparison_CET_IEOM} for various chain lengths $N$ (CET) and 
loop orders $m$ (iEoM). 
Furthermore, the short-time behavior of $g(t)$ is determined analytically by an expansion in powers of $t$, 
cf.\ \Cref{app:taylor}, and is plotted by a dashed line as reference.
The Green's function starts at $g(t\!=\!0)=\num{.5}$ because the hole creation only works if an electron with
the appropriate spin is present. Due to the assumed spin disorder this holds in $50\%$ of the cases.}

The time dependence in \Cref{img:paper2020_1D_A_comparison_CET_IEOM}  resembles a damped oscillation. 
But since we are dealing with a closed quantum system no relaxation can occur, but the superposition
of coherent oscillations is possible. In particular since we are dealing with a large mixture
of spin backgrounds it is plausible that the damping stems from strong dephasing of very many
eigenstates of the hole motion. In CET, no finite size effects appear in the studied time interval up 
to $t_\mathrm{max}=20$ (not fully shown here) as supported by the coincidence of the results
for $N=10$, $N=14$, and $N=18$. A further analysis of finite size dependence is therefore not required and 
enables us to use simulations of the largest possible system sizes in subsequent computations by CET.
The iEoM results agree very well with the CET results except for low loop order $m$. We emphasize that the iEoM dynamics consists  of oscillatory contributions exclusively; 
no contributions to $g(t)$ decrease or increase exponentially due to the guaranteed unitarity 
of the dynamic matrix \cite{Kalthoff2017,Bleicker2018}.

For the calculations based on iEoM, we use the maximum available loop order $m$ in the following. 
It varies and is strongly dependent on the topology of the lattice as well as on the number of 
physical processes considered. The number of processes depends on whether only first-order contributions 
in $\nicefrac{t_0}{U}$ with $T_0$, second-order contributions without spin-spin interaction ($H_\mathrm{0,eff}$),
 or the complete \tjm{} is considered.  {The numerically most challenging case is given by a high coordination number 
$z$ in combination with the complete \tjm{}, i.e., for 2D and $H_\mathrm{eff}$. For this case, we reached $m=3$. Larger loop order are prohibited by the required memory.}

\subsection{Determination of band edges}
\label{ss:band_edges}

 {A particularly interesting issue in the dynamics of a hole inserted into a disordered Mott insulator
is the width of the Hubbard bands. In particular, we are interested in the lower band edge 
of the upper Hubbard band. In the particle-hole symmetric case this is equivalent to the
upper band edge of the lower Hubbard band which reflects the hole motion. The necessary minimum energy eigenvalue 
$\omega_\mathrm{min}$ can be determined particularly advantageously and systematically from the iEoM results
by extrapolating $\omega_\mathrm{min}(m)$ in the loop order $m\to\infty$. We emphasize that this procedure considers the translationally invariant infinite system for any value of $m$ and takes processes into account of larger and larger
spatial range upon increasing $m$. Thus, for $m\to\infty$, the system corresponds to the entire lattice 
including all physically relevant processes. Because of the systematic nature of this expansion and 
the absence of finite-size effects, the iEoM approach is particularly appropriate for the
discussion of the band edges of the Hubbard bands and their supports. In return, we will see later
that the CET yields a better access to the overall shape of the Hubbard bands.}

\begin{figure}[ht]
  \centering
  \includegraphics[trim=32 15 45 45,clip,width=.5\textwidth]{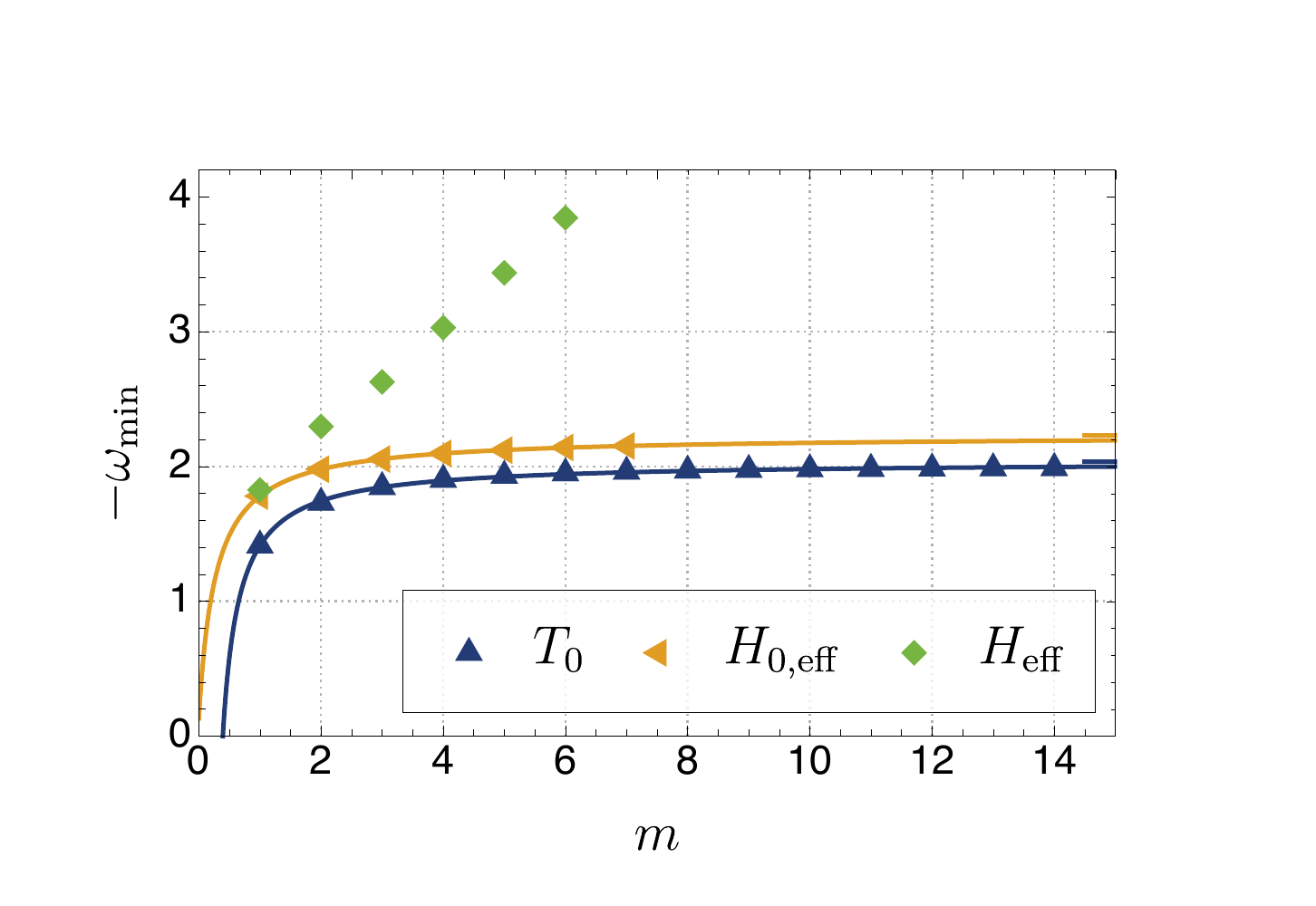}
  \caption{ {Extrapolation of the minimum eigenvalue $\omega_\mathrm{min}$ for a one-dimensional chain and 
	the parameters (A) in the loop order $m$ of the iEoM. The minimum eigenvalues for the different loop orders 
	are shown by symbols; solid lines mark fits of the form $-\omega_\mathrm{min}(m) =\nicefrac{a}{(m-b)}+c$. 
	The fit parameter $c=-\omega_\mathrm{min}(m\to\infty)$ is displayed using short 
	horizontal bars at the right boundary of the graph.}}
  \label{img:paper2020_1D_A_extrapolation_IEOM}
\end{figure}

The extrapolation is shown for the one-dimensional chain and  {parameters (A), cf.\ \eqref{eq:parameter_regimes_A};
the results for parameters (B), cf.\ \eqref{eq:parameter_regimes_B}, are qualitatively the same.
\Cref{img:paper2020_1D_A_extrapolation_IEOM} depicts the results. The different symbols mark the different cases 
depending on which processes are included in the Hamiltonian. The more processes are included in the Hamiltonian,
the lower is the  maximum achievable loop order $m$. If the minimum eigenvalues asymptotically converge towards a 
finite value}
\begin{equation}
	\label{eq:omega_min_ieom}
	c:=-\omega_\mathrm{min}(m\to\infty)	
\end{equation}
we determine this value by the  fit 
\begin{equation}
\label{eq:fit}
	-\omega_\mathrm{min}(m) = \frac{a}{(m-b)}+c.
\end{equation}
 {These fits are displayed by solid lines in \Cref{img:paper2020_1D_A_extrapolation_IEOM}; they describe the 
data shown by symbols very well. The asymptotic minimum eigenvalue $c$ is marked by a horizontal bar at the 
right boundary of the graph. Several observations are in order.
The more different couplings are included in the Hamiltonian the lower is the maximum loop order
$m$. The key observation, however, is that there is no convergence if the magnetic exchange is
included. As long as only hole hopping is considered, i.e., the Hamiltonian is $T_0$ or $H_{0,\text{eff}}$,
clear convergence can be observed and the band edge can be determined reliably by fitting $-c$.
If the magnetic exchange coupling is included, i.e., the Hamiltonian is $H_\text{eff}$, 
the convergence according to \eqref{eq:fit} is lost and the band edge diverges linearly
with $m$. This provides very strong evidence for an \emph{unbounded} support of the corresponding
spectral density of the Hubbard band.}

 {This large qualitative difference comes as a surprise. But it can be understood by analyzing 
the magnetic degrees of freedom of the Mott insulating phase right at half-filling.
This is an antiferromagnetic Heisenberg model with eigenenergies between the ground
state energy $E_\text{min}<0$ and the maximum energy for fully polarized states $E_\text{max}>0$.
Both energies are extensive, that means, they are proportional to the system size $N$
implying that they are infinite in the thermodynamic limit. Thus, the disordered spin ensemble,
which we consider as initial phase, can be expanded in eigenstates of which the eigenenergies
range from minus to plus infinity. If a hole is inserted these eigenstates are disturbed locally
at the site of the added hole. These disturbed states can again be expanded in eigenstates
of the singly-doped \tjm{}. It is highly plausible that this expansion also consists of
eigenstates with eigenenergies from minus to plus infinity. Hence, an unbounded support
for the spectral density appears naturally.}

 {We stress that the above qualitative argument does not \emph{prove} that the support is
unbounded, but it provides a plausible explanation for an unbounded support. One may object that the local
disturbance by the added hole cannot change the energy by an infinite amount. But this argument
only refers to the expectation value of the energy before and after the insertion of the hole.
The above argument does not make statement on the matrix elements of the transitions so that
a finite change of the energy expectation is perfectly consistent with the infinite support.}

 {As an illustration that similar scenarios
exist we refer to the example of spectral densities of local Green's functions
of impurities in metallic hosts. Here the disturbance is also local, but the support
of the spectral density is defined by all possible transitions from $\vec k$ to $\vec k'$ so
that the support generically is as large as the full band width. 
We will come back to the shape of the spectral density of the hole motion in the disordered
spin background below.}

\section{Results for the Chain}
\label{s:results-1D}

 {We consider explicit results for the local spectral densities and their lower band edge
if it is finite. If no band edge exists we study the tails of the spectral densities.}

\subsection{Spectral densities}
\label{ss:spectral_densities_1D}

\begin{figure}[ht]
  \centering
  \includegraphics[trim=10 15 45 45,clip,width=.5\textwidth]{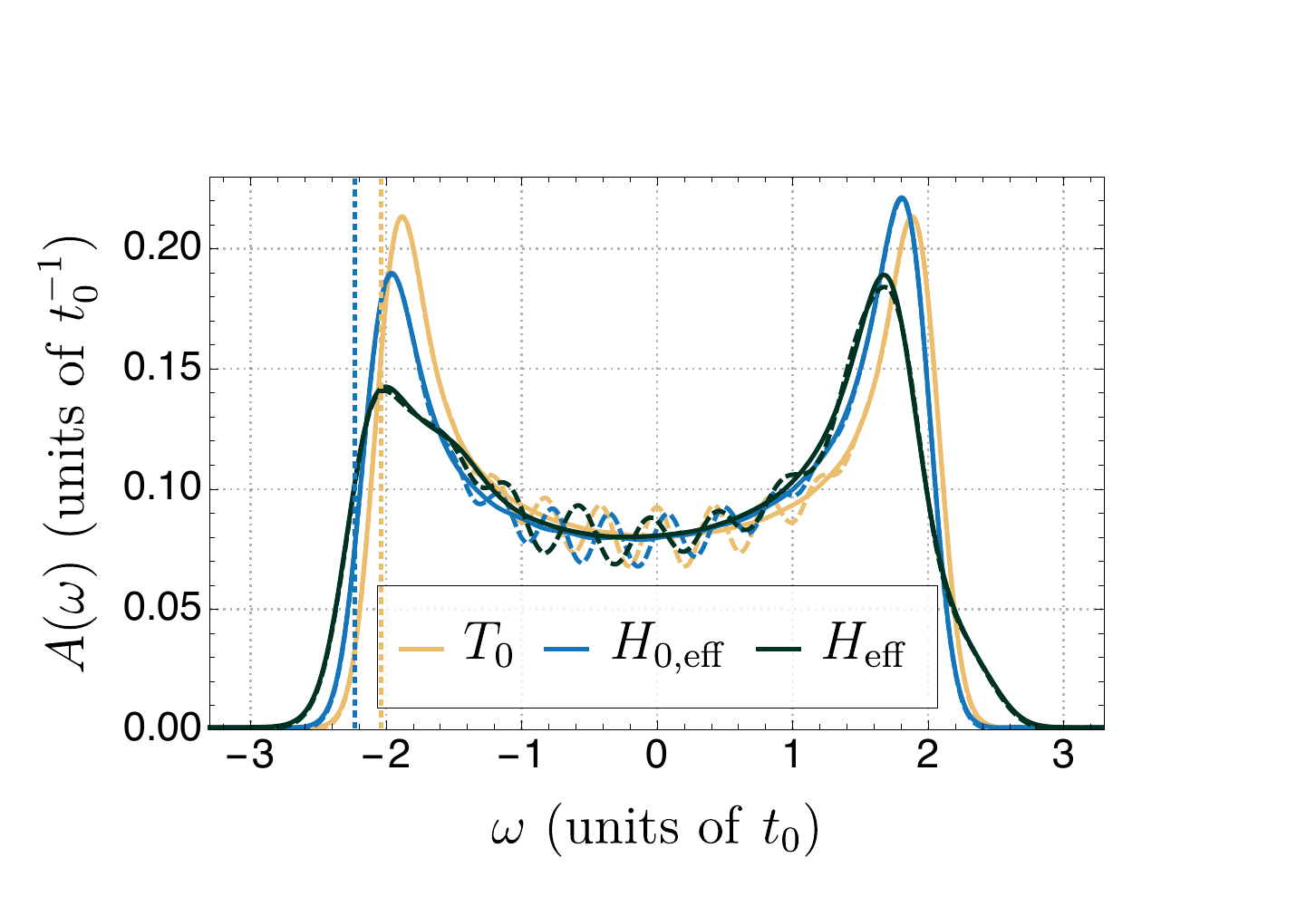}
  \caption{Spectral density $A(\omega)$ vs.\ $\omega$ for a one-dimensional chain and the parameter set (A) in
	\eqref{eq:parameter_regimes_A}, artificially broadened by $\sigma=0.15t_0$. 
	Solid lines represent CET results, dashed lines iEoM results. 
	The band edges $\omega_\mathrm{min}$ determined from \eqref{eq:omega_min_ieom} are indicated by vertical dashed lines.}
  \label{img:paper2020_1D_A_sigma_0.15}
\end{figure}

\begin{figure}[ht]
\vspace{-.5cm}
  \centering
  \includegraphics[trim=10 15 45 45,clip,width=.5\textwidth]{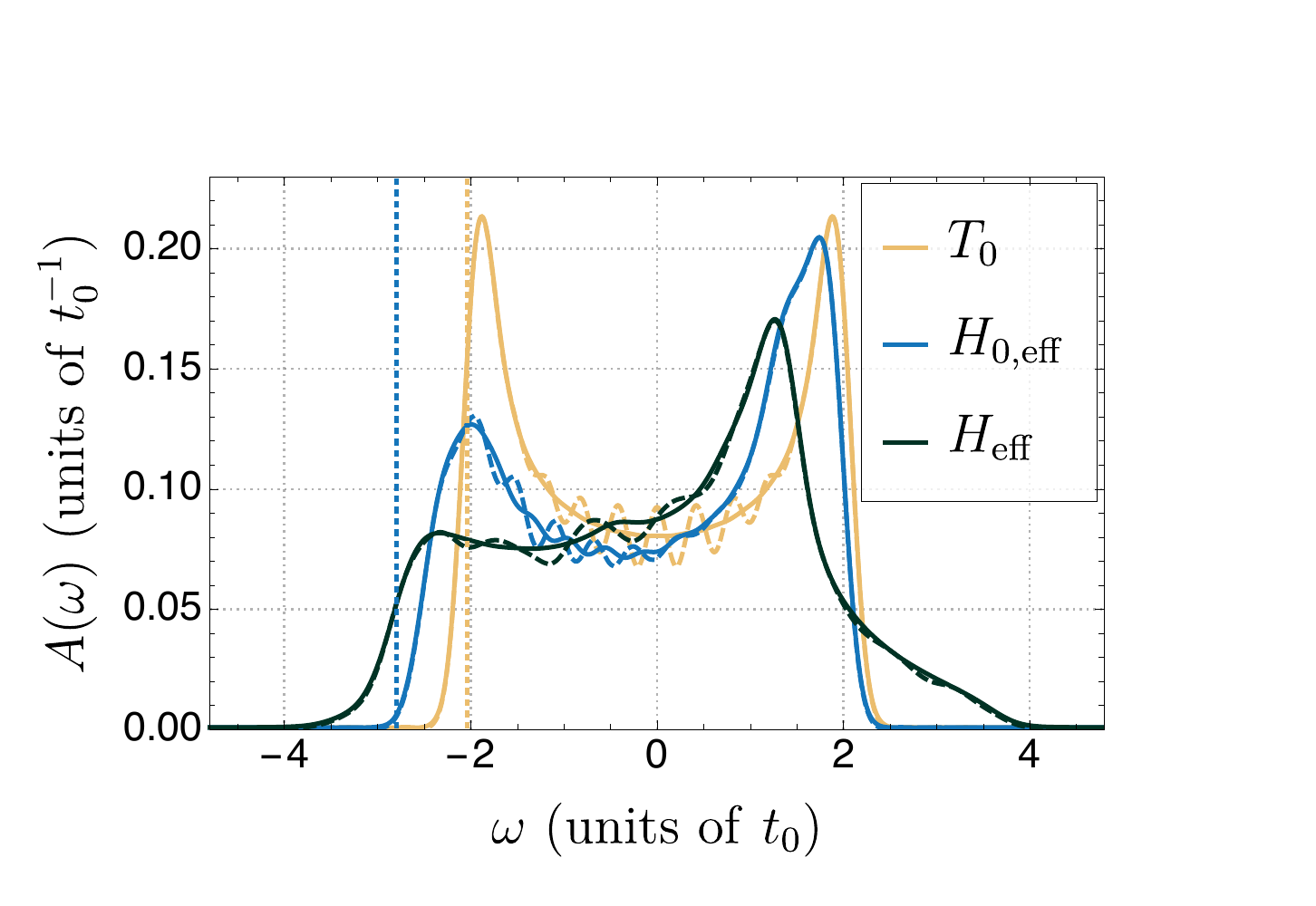}
  \caption{Same as \Cref{img:paper2020_1D_A_sigma_0.15}, but for the parameters (B) in
	\eqref{eq:parameter_regimes_B}.}
  \label{img:paper2020_1D_B_sigma_0.15}
\end{figure}

The spectral densities $A(\omega)$ for the one-dimensional chain for the three cases $T_0$, $H_{0,\text{eff}}$,
 and $H_\text{eff}$ are displayed in \Cref{img:paper2020_1D_A_sigma_0.15,img:paper2020_1D_B_sigma_0.15}.
For $H_{0,\text{eff}}$ and $H_\text{eff}$ the difference of two parameter sets (A) and (B) matters.
The results from the different methods used, CET (solid) and iEoM (dashed), agree very well in all cases. 
Note that the data has been broadened by $\sigma=0.15t_0$.

Both the upwards and the downwards flanks of the spectral density, as well as the characteristic shape including 
the peak positions are accurately reproduced. The wiggling of the iEoM results around 
$\omega=0$ results from a few discrete, Gaussian broadened peaks. Higher loop order $m$ and thus increased basis
would lead to smoother spectral densities. 

 {The spectral density $A(\omega)$ for $T_0$ is symmetric about $\omega=0$. This is expected because
it is obvious from $T_0$ that it corresponds to nearest-neighbor hopping which implies 
symmetric local densities-of-states (DOS). This has been shown rigorously in the 1D case
in the limit $U\to\infty$ \cite{mielk91aa,kumar09}. This also explains the value
of the lower band edge $\omega_\mathrm{min}=\num{-2}t_0$ which our extrapolation
reproduces within a relative error of $2\%$. We emphasize that the 
determination of the band edges does not involve any broadening. 
 The pronounced peaks are the van Hove singularities
which are smeared out by finite-size effects or finite loop order and the artificial broadening.
Otherwise, they would show up as $1/\sqrt{\Delta \omega}$ divergences.
In fact, the analytical results \cite{mielk91aa,kumar09} imply that the DOS is given by
\begin{equation}
A(\omega) = \frac{1}{2\pi}\frac{1}{\sqrt{\omega^2-4t_0^2}}.
\end{equation}}

 {If the spin-dependent and spin-independent hopping is included, i.e., we consider
$H_{0,\text{eff}}$, the support of the spectrum increases. For parameter
set (A) by about $10\%$ and for set (B) by almost $20\%$. Since the DOS satisfies 
the sum rule $\int A(\omega)\dd{\omega}=\nicefrac{1}{2}$, a larger support 
 necessarily translates into a reduced average height. In addition, one
clearly sees that the DOS loses its symmetry: the left van Hove peak becomes lower
than the right one. A physical explanation for this behavior is left to future research.}

 {If the magnetic exchange, i.e., the spin-spin interaction, is included as well we consider
the dynamics induced by $H_\text{eff}$. The corresponding data is shown by the darkest 
curves in \Cref{img:paper2020_1D_A_sigma_0.15,img:paper2020_1D_B_sigma_0.15}. 
The broadened curves show a larger asymmetry between the left and the right peak
compared to the curves for $H_{0,\text{eff}}$. For parameter set (B) the left
peak is reduced to only a shoulder. In addition, its seems that
the band edges have been slightly more shifted and broadened.
But from the previous analysis of the non-convergence of the band edge we know that
this impression is misleading. In fact, there is no finite support of the DOS
anymore. We will analyze the tails of the DOS quantitatively in the next section.}

\subsection{Gaussian tails}
\label{ss:tails_1D}

 {In Sect.\ \ref{ss:band_edges} we already found striking evidence that the spectral density 
differs qualitatively if the magnetic exchange is considered or not. Here we come back to
this point and study the case with magnetic exchange, i.e., $H_\text{eff}$, in more detail.
We want to find out what the tails of the spectral densities look like.
Motivating starting point is the fact that the orientation of each spin at half-filling
is chosen randomly and independently for each site in the completely disordered
spin ensemble. Hence, an infinite number of independent random processes influences
the matrix elements entering the spectral densities and
their tails in particular. The central limit theorem suggests that
the resulting tails are of Gaussians nature. This is consistent with the
finding that the support of the spectral densities is infinite.
But we emphasize that the hypothesis of Gaussian tails represents an educated guess
at this stage. Therefore, we put this hypothesis to a quantitative test.}

\begin{figure}[ht]
  \centering
  \includegraphics[trim=10 15 45 45,clip,width=.5\textwidth]{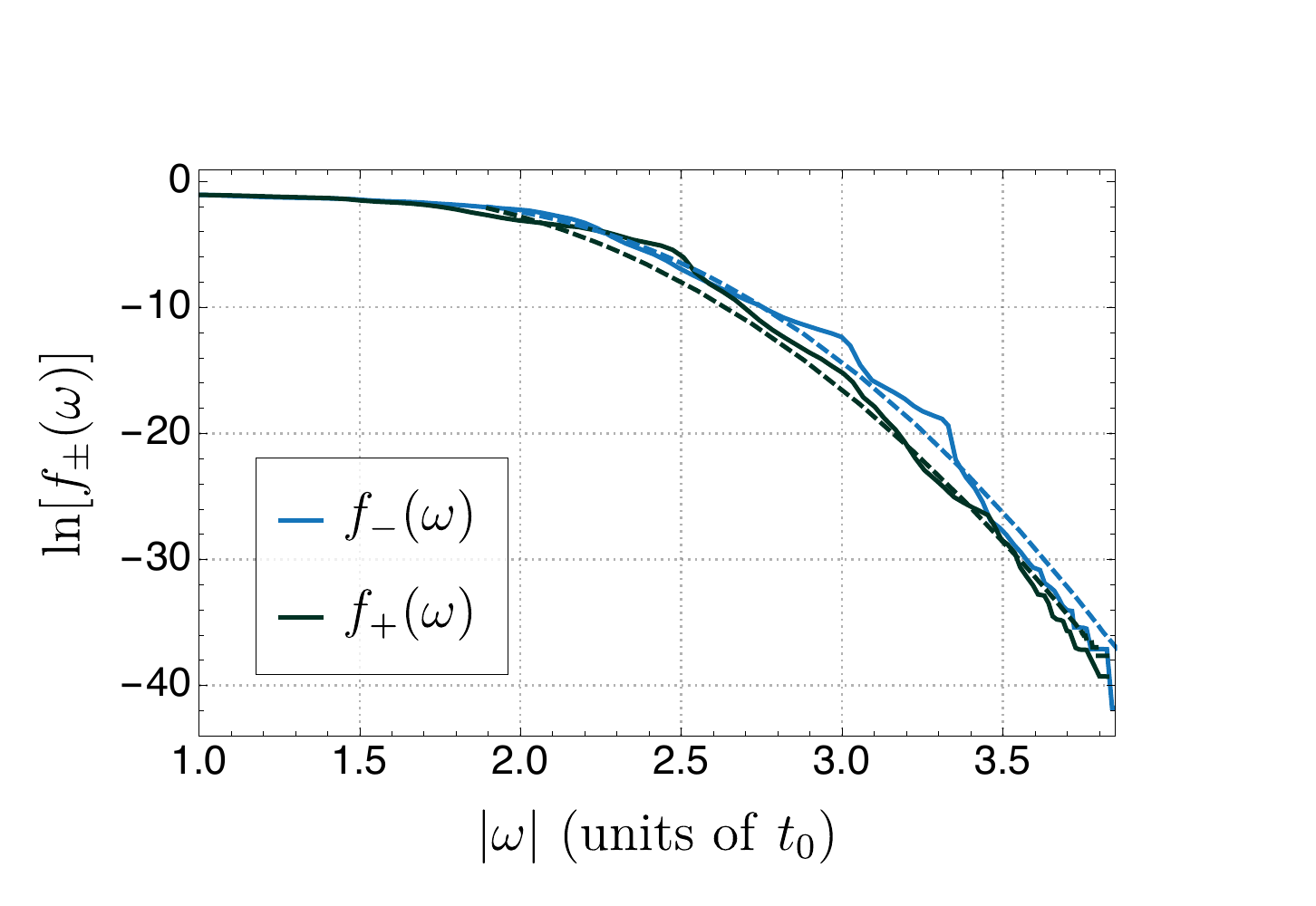}
  \caption{ {Analysis of the lower and the upper tail of the spectral density as obtained 	by iEoM via the logarithm of the primitives defined in Eq.\ \eqref{eq:primitives}
	for the parameter set (A). The primitives are shown as solid lines;
	the fits defined in Eq.\ \eqref{eq:fit_gauss} are shown as dashed lines. 
	The optimum fit parameters read
	$W_-=0.28$, $\sigma_-=0.24t_0$, $x_-=-1.88t_0$ for the lower tail and 
	$W_+=0.80$, $\sigma_+=0.26t_0$, $x_+=1.62t_0$ for the upper tail.}}
  \label{img:paper2020_1D_A_ieom_tails}
\end{figure}

\begin{figure}[ht]
  \centering
  \includegraphics[trim=10 15 45 45,clip,width=.5\textwidth]{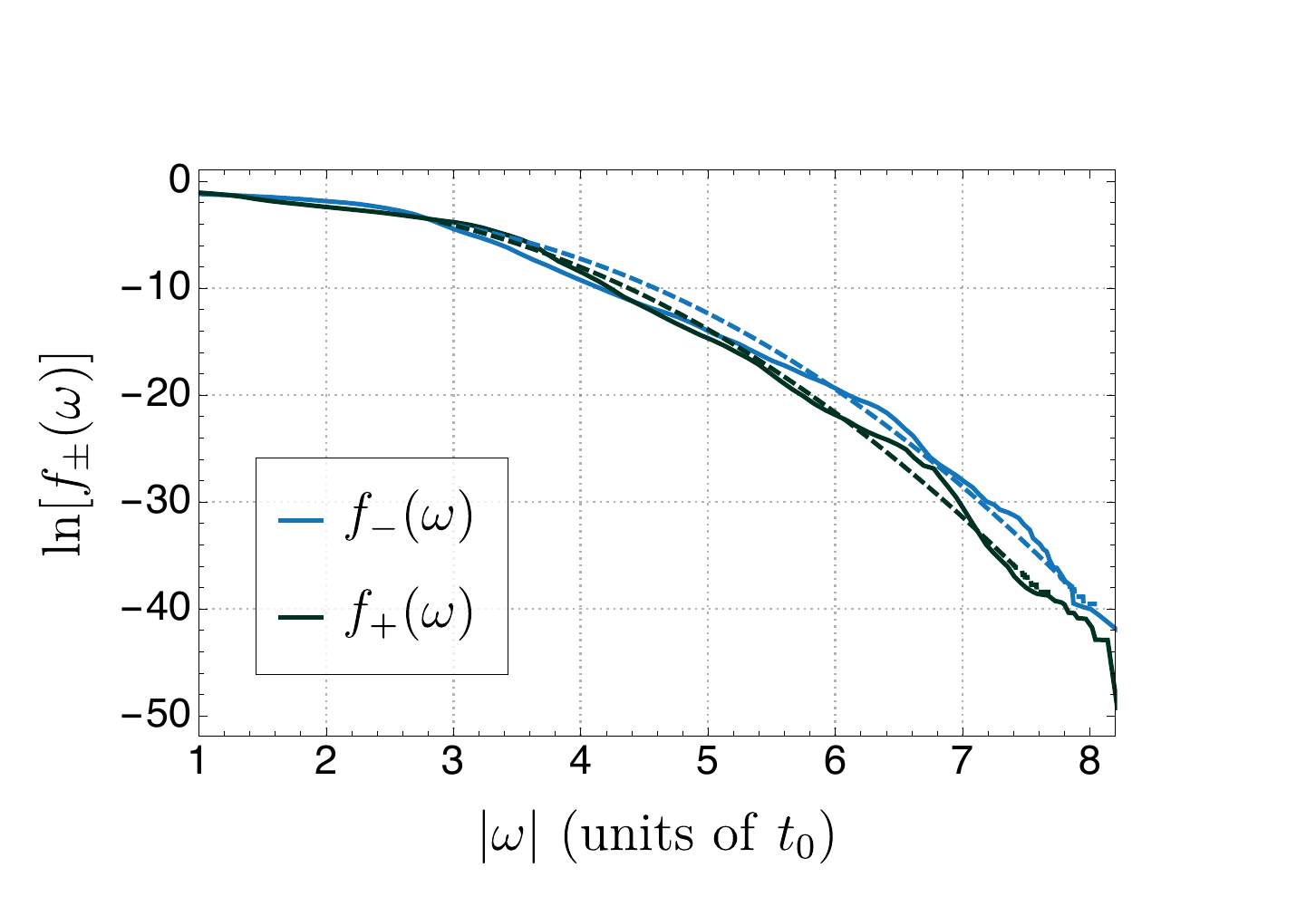}
  \caption{ {Same as \Cref{img:paper2020_1D_A_ieom_tails}, but for parameters (B) and the fit parameters 
	$W_-=0.12$, $\sigma_-=0.69t_0$, $x_-=-2.26t_0$ for the lower tail and 
	$W_+=0.37$, $\sigma_+=0.70t_0$, $x_+=1.81t_0$
	for the upper tail.}}
  \label{img:paper2020_1D_B_ieom_tails}
  \vspace{-.5cm}
\end{figure}

For this test we have to refrain from using any broadening because this
induces artificial tails which conceal the intrinsic physics. Thus we do
not consider the spectral density itself but its primitive as is routinely
done in probability theory. We define
\bes
\label{eq:primitives}
\begin{align}
f_-(\omega) &= \int_{-\infty}^\omega A(x)\dd{x}
\\
f_+(\omega) &= \int_{\omega}^\infty A(x)\dd{x},
\end{align}
\ees
where $f_-$ is used to study the lower tail $\omega\to-\infty$ and $f_+$ for the upper tail $\omega\to\infty$. If the tails are Gaussian we have 
\bes
\label{eq:fit_gauss}
\begin{align}
f_-(\omega) &\approx \frac{W_-}{\sqrt{2\pi}\sigma_-}\int_{-\infty}^\omega \exp(-(x-x_-)^2/(2\sigma_-^2))\dd{x}
\\
 &= \frac{W_-}{2}\left(\erf(\omega_-)+1\right)
\\
f_+(\omega) &\approx \frac{W_+}{\sqrt{2\pi}\sigma_+}\int_\omega^\infty \exp(-(x-x_+)^2/(2\sigma_+^2))\dd{x}
\\
 & = \frac{W_+}{2}\left(1-\erf(\omega_+)\right),
\end{align}
\ees
where 
\bes
\begin{align}
\omega_- &:= (\omega-x_-)/(\sqrt{2}\sigma_-)
\\
\omega_+ &:= (\omega-x_+)/(\sqrt{2}\sigma_+).
\end{align}
\ees
Note that three free parameters need to be determined by fitting. 
Since exponentially small
values occur we plot $\ln(f_\pm)$ as function of $|\omega|$ in Fig.\ 
\ref{img:paper2020_1D_A_ieom_tails}
and compare it with the fits \eqref{eq:fit_gauss}. The best fit parameters are
given in the caption. The corresponding results for the parameter set (B) can be found in 
\Cref{img:paper2020_1D_A_ieom_tails,img:paper2020_1D_B_ieom_tails}. 

The agreement between the data obtained by iEoM and the fits
is very good. The logarithm of the 
iEoM data clearly shows roughly parabolic shape
consistent with Gaussian tails. Of course, some fluctuations 
around the rigorous error functions occur. But we stress that
the agreement found for all four fit extends over 15 (!) orders
of magnitude if one converts the differences on the
log-scale to decimal ratios. We take this observation as strong
support for our claim of Gaussian tails.

%%% %Bethe ansatz results ... discuss 
The above finding of an infinite support of the DOS and of its Gaussian tails
is in stark contrast to the findings of Ejima and co-workers \cite{Ejima2006}
who studied the 1D Hubbard model by Bethe ansatz under the assumption of
a completely disordered spin background. This appears indeed very similar to
the physical situation studied in the present article. Ejima \emph{et al.}
determine a critical $\Uc$ below which the assumed Mott insulating phase
becomes unstable. This implies that the Hubbard bands have finite, well-defined
band edges which vanish if the shift by $U/2$ becomes to small. This is 
at variance with the above findings.

Two explanations for this difference are conceivable. First, Ejima \emph{et al.}
study the Hubbard model as such without prior mapping to the \tjm{}. This mapping certainly
influences matrix elements and hence it will have a certain effect on the shape of the spectral
density of hole motion. Yet, we think it is unlikely that this mapping changes a finite
support to an infinite support, i.e.,  it is not plausible that matrix elements
between eigenstates strongly differing in energy are induced by this mapping which 
are strictly zero in the Hubbard model itself. At present, however, we cannot exclude
this explanation.

Second, the assumption of a totally disordered spin background is
physically subtle. It is not difficult to construct the ensemble. But it
must be kept in mind that it does not constitute a physically stable equilibrium
situation except in the limit $J\ll T \ll t_0$ which represents an extreme
parameter region with very large $U$ (recall $J=4t_0^2/U$). Hence, the occurrence
of large energies in the spectral density of hole motion, induced by the
large energy differences of the magnetic background
in $H_J$,  appears plausible. We presume that the energy differences in the
magnetic background are not included in the way the Bethe ansatz approach
to hole hopping in disordered spins is conducted. But this interesting
issue certainly calls for further elucidation.

The  analyses in two dimensions analogous to the above analyses for the chain
are not conclusive currently because of the limit
loop order $m$ that can be reached. But the preliminary results 
point into the same direction as in one dimension. 
In view of the conceptual interest of this issue
 a follow-up study should  expand on this.

\section{Results for the square lattice}
\label{s:results-2D}

Analogous to calculations for the chain, spectral densities and band gaps can also be determined on the square lattice. We emphasize that such a calculation is not merely an enlargement of the dimension, but 
introduces additional physical processes. For instance, there are four nearest neighbors on a square lattice
 instead of two nearest neighbors on the chain yielding a more densely populated Hamiltonian matrix. In parallel,
for the same tractable cluster size $N$, only $\sqrt{N}$ hopping processes are available for NN hopping 
and correspondingly fewer for NNN or 3NN hopping until wrap-around effects occur in two dimensions.
Thus, describing the dynamics without finite-size effects becomes immensely more demanding. As a result, 
the obtained densities are not as smooth as in one dimension and show more wiggling.
The iEoM treats the thermodynamic limit by construction, but it cannot reach
the same accuracy as in one dimension either because the additional physical processes
reduce the maximum loop order $m$ that can be reached.

\begin{figure}[ht]
  \centering
  \includegraphics[trim=10 15 45 45,clip,width=.5\textwidth]{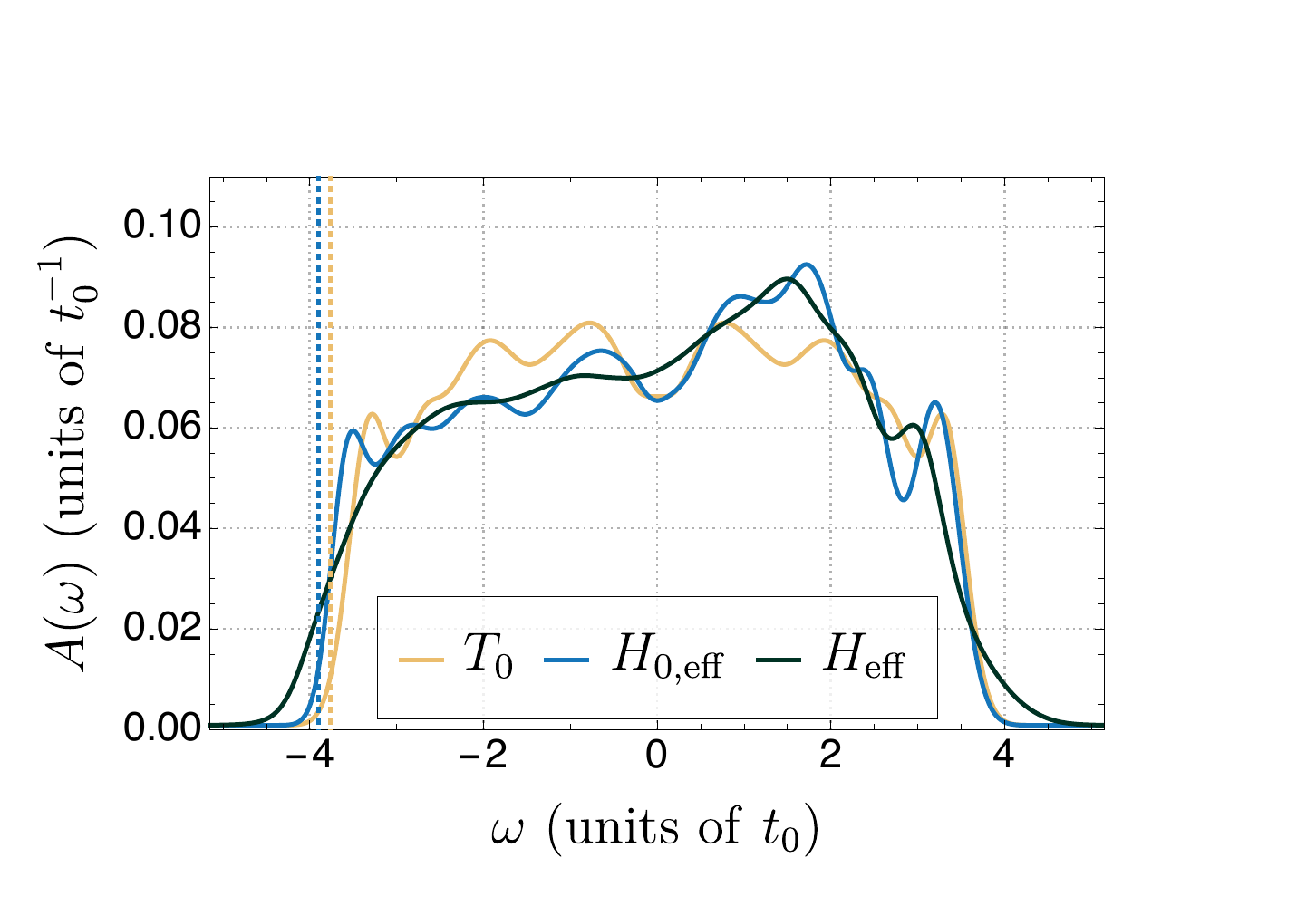}
  \caption{Spectral density $A(\omega)$ for hole motion on the square lattice and parameter
	set (A). See caption of \Cref{img:paper2020_1D_A_sigma_0.15} for further explanations.}
  \label{img:paper2020_2D_A_sigma_0.15_CET}
\end{figure}

\begin{figure}[ht]
  \centering
  \includegraphics[trim=10 15 45 45,clip,width=.5\textwidth]{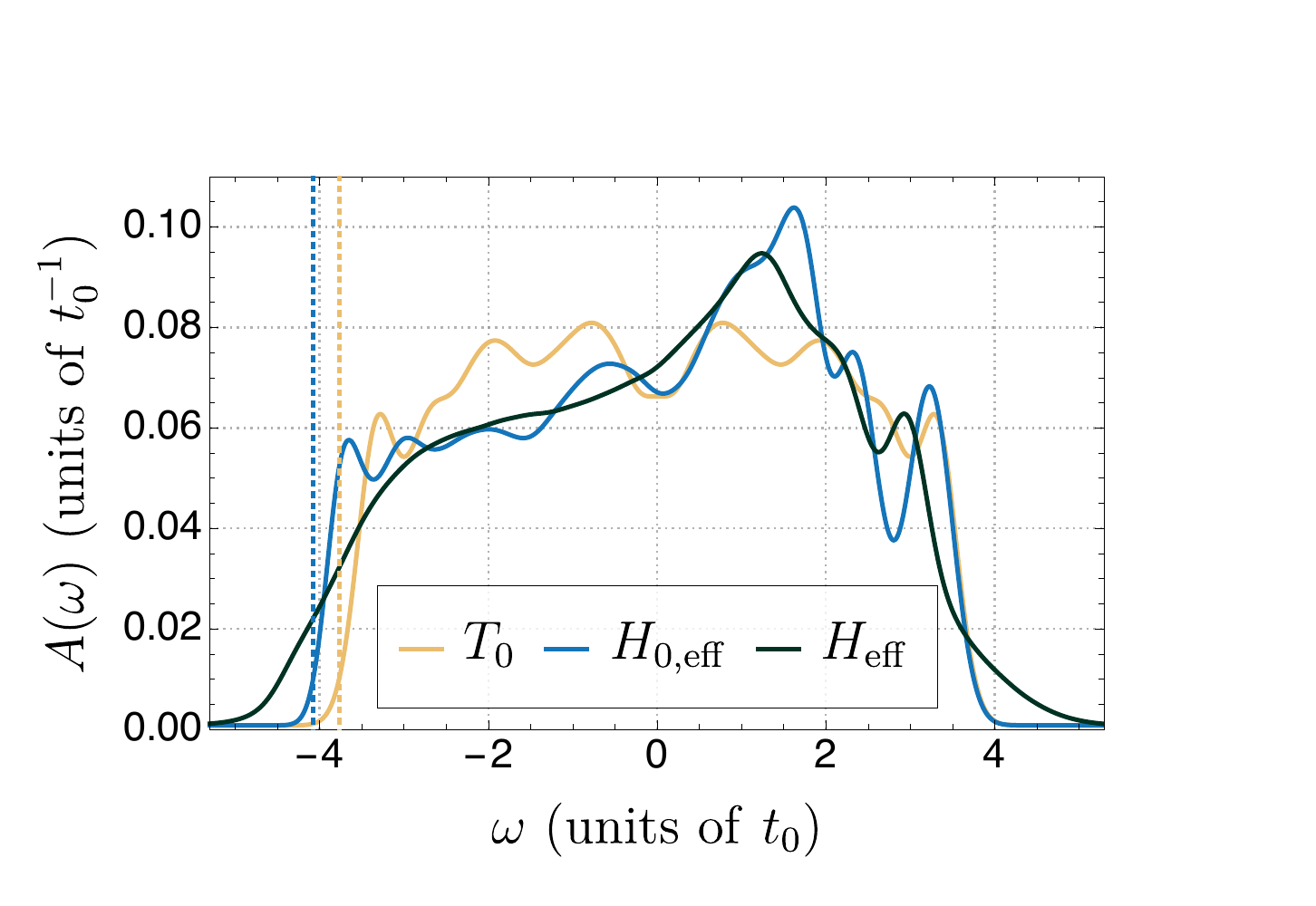}
  \caption{Spectral density $A(\omega)$ for hole motion on the square lattice and parameter
	set (B). See caption of \Cref{img:paper2020_1D_A_sigma_0.15} for further explanations.}
  \label{img:paper2020_2D_B_sigma_0.15_CET}
\end{figure}

In order to achieve a higher number of hopping processes before wrap-around effects kick in, we resort to a trick and rotate the studied square cluster by \SI{45}{\degree}. 
Then its edge length is given by $\sqrt{2}n$ according to Pythagoras where $n$ is
the number of vertical and horizontal NN steps to pass from one corner of the square cluster to the adjacent one. Thus, the total number of sites is $N=2n^2$. For $n=3$ 
we have to treat 18 sites which is still feasible. The advantage is that a 
wrap-around only occurs after $2n=6$ NN hops. We emphasize that a naive choice of 
the square cluster with an equal number of hops for wrap-around would have required 
$N=6^2=36$ sites. The corresponding Hilbert space would be almost $\num{2.6e5}$ 
times larger.

The overall shape of the spectrum is significantly altered compared to the case
of a chain, see Figs.\ \ref{img:paper2020_2D_A_sigma_0.15_CET} and 
\ref{img:paper2020_2D_B_sigma_0.15_CET}
for CET results. Results obtained by iEoM are presented and compared to CET results for larger broadening  in \Cref{app:2d}. 
In contrast to the two distinct  van Hove singularities in the 
DOS the 2D case reveals a spectral density of approximately elliptical to 
rectangular shape.
It is symmetric if only $T_0$ is considered and becomes asymmetric as soon
as the Hamiltonian is extended in agreement with what we found in one dimension.
Note, however, that the lower band edge for $T_0$ is not $-4t_0$, but in its vicinity (see vertical dashed lines in \Cref{img:paper2020_2D_A_sigma_0.15_CET,img:paper2020_2D_B_sigma_0.15_CET}), as one
would have expected for simple NN hopping in contrast to the 1D case
where we found $-2t_0$ in accordance with analytical arguments \cite{mielk91aa,kumar09}.
The reason is that in 1D at $U=\infty$ perfect spin-charge separation for NN hopping
occurs, i.e., the sequence of spins is not changed at all by the hole motion.
On the square lattice, this is no longer true since loops occur and only 
Trugman paths \cite{trugm88} allow for hole motion without changes of the spin order.

For the square lattice, the semi-analytically determined band edges 
$\omega_\mathrm{min}$ 
are significantly closer to each other for the two cases displayed than for the one-dimensional case. For the parameter set (A) we attribute this to the 
altered dimensionality.
For parameter set (B), this effect is enhanced by the smaller value of 
the exchange coupling $J$, i.e. because of 
$J_\mathrm{1D,B}=1>\nicefrac{1}{2}=J_\mathrm{2D,B}$. We emphasize that our
results agree with results of previous research, for instance $\omega_c=\num{-4.4}t_0$ for the full \tjm{} as given in \refcite{Reischl2004}. This is in the range of the left flanks where the DOS starts rising significantly, 
cf.\,black curves in \Cref{img:paper2020_2D_A_sigma_0.15_CET,img:paper2020_2D_B_sigma_0.15_CET}. An exact determination of 
the band edge is not possible due to the previously motivated Gaussian tails. 
We stress that this finding is not an artefact of the iEoM technique, 
but reflects the underlying physics.

Analogous to the one-dimensional case, a broadening of the spectrum upon including 
more and more processes is observed on the square lattice. For parameter set (A) 
the spectrum broadens from  $T_0$ to the complete \tjm{} $H_\mathrm{eff}$ by about 
\SI{13}{\percent}; for parameter set (B) by almost \SI{25}{\percent}. Instead of 
peaks at the boundaries of the DOS one observes knee-like flanks.

\section{Summary and Conclusions}
\label{s:summary}

In this work, we studied the dynamics of single hole in a disordered spin background
for the \tjm{} as it results for the Mott insulating phase from the Fermi-Hubbard model
by an expansion in  $\nicefrac{t_0}{U}$ where $U$ is the local repulsion and $t_0$ the
nearest-neighbor (NN) hopping.
For this purpose, we systematically extended the NN hopping $T_0$ via 
spin-dependent and spin-independent NNN and 3NN hopping to the full \tjm{} including 
the spin-spin exchange interaction $\vec{S}_i\vec{S}_j$.
For the one-dimensional chain and the two-dimensional square lattice
we computed the lower band edges of the Hubbard bands and the shape of the local spectral
density, i.e., the density-of-states (DOS). This is achieved by two approaches,
the iterated equations of motion (iEoM) and the Chebyshev expansion technique (CET).

The CET is a well-established numerically exact method for the analysis of finite clusters
whose effort increases exponentially with the Hilbert space size.
The iEoM addresses the infinite translationally invariant lattice, i.e.,
the thermodynamic limit. The systematic enlargement of the iEoM to processes 
of larger and larger spatial range by increasing the loop order $m$ 
renders profound statements on the existence and the value of well-defined band
edges possible. We found strong evidence that the support of the DOS is only finite
if hole hoppings enter the Hamiltonian exclusively. Once magnetic exchange
is switched on the support becomes infinite and the DOS develops Gaussian tails.
This effect has not yet been observed or discussed in the literature to our knowledge.
In contrast, a previous analysis of the Hubbard model based on Bethe ansatz
found finite band edges for the hole motion in a disordered spin background 
\cite{Ejima2006}. At present, it is unclear whether this difference results from 
the study of the different, though related models, Fermi-Hubbard model and \tjm{}, 
or from differences in the treatment of the magnetic dynamics and surely
merits further investigation.

Our analysis has become possible by the use of the iEoM. The evidence for Gaussian
tails is rather stringent in one dimension, but
indications for Gaussian tails exist as well  in two dimensions.
The substantially higher numerical effort in two dimensions 
calls for further efforts to corroborate the advocated scenario further.
Analogous studies for other lattices in two dimensions and also in three dimensions to 
study the influence of lattice topology are conceivable and desirable.\\\\

\begin{acknowledgments} 
We gratefully acknowledge financial support by the Konrad Adenauer Foundation (PB) as 
well as by the German Research Foundation (DFG) in projects UH 90-13/1 (GSU) and UH 90-14/1 (DBH) as well as in project B9 of ICRC 160 (GSU) together
with the Russian Foundation for Basic Research. 
All authors contributed equally to this work, PB and GSU wrote the manuscript. 
The authors are indepted to Florian Gebhard for helpful and fruitful discussions.
\end{acknowledgments}

%
%% EMBEDDED_BIBLIOGRAPHY_POSITION

%\newpage
% \begin{widetext}
\begin{appendix} 

\section{Approximation of the short-time behavior of $g(t)$}
\label{app:taylor}

The behavior of the retarded Green's function $g(t)$ for $t\gtrsim0$ 
can be estimated analytically by an expansion in powers of $t$.
The result reads
\begin{equation}
\label{eq:taylor}
g(t)\approx - \frac{i}{2} \left( 1  + 
\Braket{\left[H, \lup{i}(0) \right] \left[H, \ltup{i}(0) \right] } t^2 \right) + \mathcal{O}\left(t^3\right)
\end{equation}
where the translational invariance in time, i.e.,
\begin{equation}
  g'(t)=g'(-t)=\Braket{\left[H, \lup{i}(0) \right] \ltup{i}(-t) }
\end{equation}
allows us to apply the second derivative to the second operator
\begin{equation}
  g''(t)=\dv{g'(-t)}{t}=-i\Braket{\left[H, \lup{i}(0) \right] \left[H, \ltup{i}(0) \right] }.
\end{equation}
In this way, a double commutator is avoided.

For clarity, we apply formula \eqref{eq:taylor} to the one-dimensional chain. The commutators appearing are
\begin{subequations}
\label{eq:app_comms}
\begin{alignat}{3}
  \left[ T_0,\lup{i} \right] &= t_0 \ldown{i\pm1}\splus{i} + \frac{1}{2}t_0 \lup{i\pm1}\sz{i} + 
	\frac{1}{2}t_0 \lup{i\pm1} \\
  \left[ T_0'',\lup{i} \right] &= t'' \ldown{i\pm2}\splus{i} + \frac{1}{2}t'' \lup{i\pm2}\sz{i} + 
	\frac{1}{2}t'' \lup{i\pm2} \\
  \left[ T_{s,0}'',\lup{i} \right] &= \frac{1}{2}t''_s \ldown{i\pm2}\splus{i\pm1} + 
	\frac{1}{2}t''_s \ldown{i\pm2}\splus{i\pm1} \sz{i} \\
&+ t''_s \lup{i\pm2} \sminus{i\pm1} \splus{i} 
 + \frac{1}{4} t''_s \lup{i\pm2} \sz{i\pm1} \\
 &+  \frac{1}{4} t''_s \lup{i\pm2} \sz{i\pm1}\sz{i}
-\frac{1}{2} t''_s \ldown{i\pm2} \sz{i\pm1}\splus{i} \\
\left[ H_J,\lup{i} \right] &= \frac{1}{4} J \lup{i} \sz{i\pm1} + \frac{1}{2} \ldown{i} \splus{i\pm1}. 
\label{eq:app_comms_J}
\end{alignat}
\end{subequations}
The remaining commutators for the case $\ltup{i}$ result from the relations 
\eqref{eq:app_comms} 
substituting  $\lup{i}\rightarrow -\ltup{i}$ as well as 
$\splus\leftrightarrow\sminus$. 
The expectation values occurring in \eqref{eq:taylor} can be calculated 
straightforwardly 
since they are to be determined at $t=0$. The trace is computed over 
states at half-filling without a hole. 
For demonstration purposes, we give the results for the expectation values that arise from
$H_J$, see \eqref{eq:app_comms_J}, 
\begin{subequations}
\label{eq:app_expvals}
\begin{alignat}{3}
\Braket{\lup{i} \sz{i\pm1}\ltup{i} \sz{i\pm1}} &= 2\cdot\frac{1}{2} \\
\Braket{\ldown{i} \splus{i\pm1} \ltdown{i} \sminus{i\pm1}} &= 2\cdot\frac{1}{4}.
\end{alignat}
\end{subequations}
Here, the first factor results from the double occurrence of the expectation value, 
once for $i+1$ and once for $i~-~1$. The expectation values from the other contributions 
can be calculated similarly. Substituting all expectation values and \eqref{eq:app_comms} 
into \eqref{eq:taylor} then yields the explicit expansion
\begin{equation}
  g(t)=-\frac{i}{2}\left[1- \left(t_0^2+t''^2+\frac{6}{16}t_s''^2+\frac{3}{32}J^2 \right)t^2\right]
	+\mathcal{O}(t^3).
\end{equation}

\section{2D results from iEoM and CET}
\label{app:2d}

In addition to the results obtained for the square lattice using CET,
convolved with $\sigma\!=\!\num{.15}t_0$, and shown in 
\Cref{img:paper2020_2D_A_sigma_0.15_CET,img:paper2020_2D_B_sigma_0.15_CET}, 
the analogous results can also be obtained using iEoM. Due to the limited
loop order $m$ they need to be broadened more strongly by Gaussians.

\begin{figure}[ht]
  \centering
  \includegraphics[trim=10 15 45 45,clip,width=.5\textwidth]{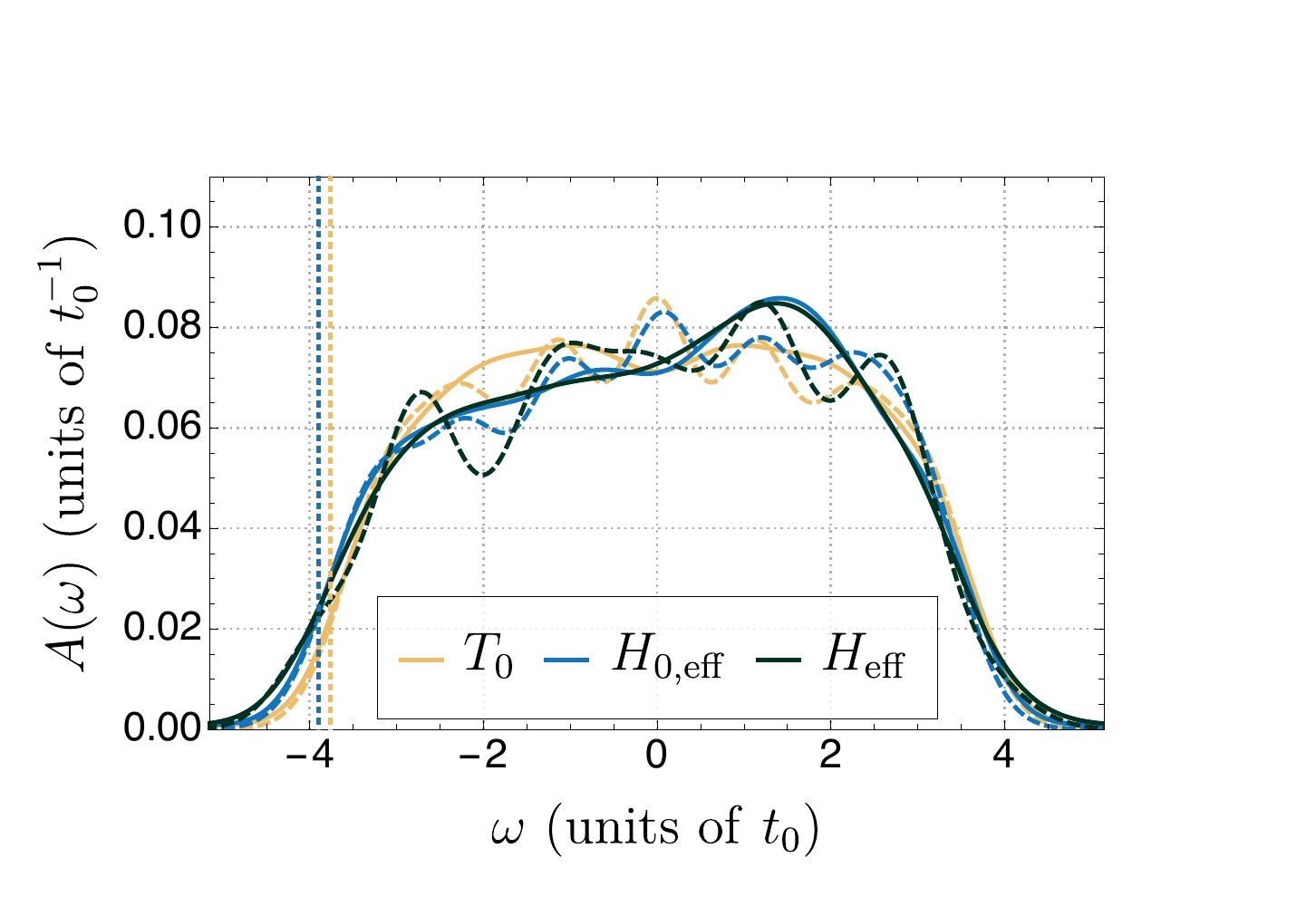}
  \caption{Spectral densities $A(\omega)$ for the square lattice, 
	parameter set (A), and the various contributions, calculated with CET 
	for $N=18$ 	(solid) and iEoM for $m=4$ for $T_0$ and $H_{0,\text{eff}}$
	and $m=3$ for $H_{\text{eff}}$ (dashed). 
	All data are convolved with $\sigma=\num{.45}t_0$. The band edges
	$\omega_\mathrm{min}$ calculated according to \eqref{eq:omega_min_ieom} 
	are indicated by vertical dashed lines.}
  \label{app:paper2020_2D_A_sigma_0.45}
  \vspace{-.35cm}
\end{figure}

\begin{figure}[ht]
  \centering
  \includegraphics[trim=10 15 45 45,clip,width=.5\textwidth]{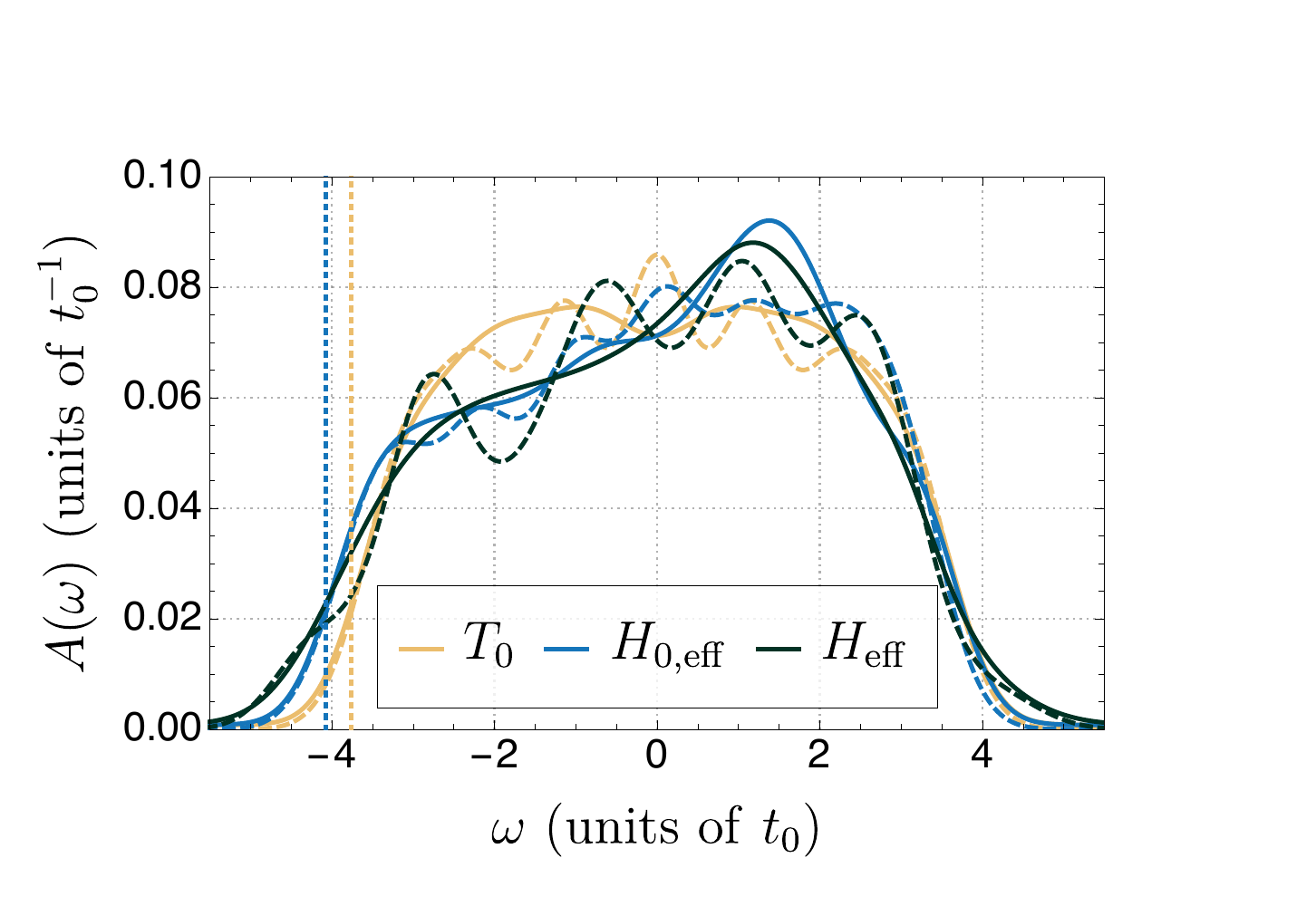}
  \caption{Same as in \Cref{app:paper2020_2D_A_sigma_0.45}, but for parameter set (B).}
  \label{app:paper2020_2D_B_sigma_0.45}
  %\vspace{-.5cm}
\end{figure}

In view of the fact that the maximum possible loop order $m$ is comparatively limited,  wiggly 
spectral densities occur. In order to ensure a reasonable comparability to CET results and 
showing  the good agreement of both methods a convolution of (all) results 
with an increased $\sigma=\num{.45}t_0$ is performed. Still, the iEoM results
display some spurious wiggles. The corresponding results for the sets
 (A) and (B) are depicted in 
\Cref{app:paper2020_2D_A_sigma_0.45,app:paper2020_2D_B_sigma_0.45}. The increased width
of the CET results compared to the ones in  
\Cref{img:paper2020_2D_A_sigma_0.15_CET,img:paper2020_2D_B_sigma_0.15_CET}
is an artefact due to the enhanced broadening. Obviously, the band edges obtained from 
the minimum eigenvalues of the Liouville matrix in iEoM 
are identical regardless of the additional broadening. 

The high degree of agreement between the two methods in the margins of the spectral density can be understood in particular on the basis of the fact that the Lanczos algorithm used is particularly accurate in the range of extremal eigenvalues. Increasing Krylov space dimensions $f$ as well as an increase of the loop order $m$ lead to an even higher similarity of the results of both methods.

\end{appendix}
% \end{widetext}

\end{document}